\documentclass[preprint,aps ,nofootinbib]{revtex4}
\usepackage[colorlinks=true,linkcolor=blue,urlcolor=blue,filecolor=black,citecolor=red,pdfstartview=FitV,pdftitle={},pdfsubject={},pdfkeywords={},pdfpagemode=None,bookmarksopen=true]{hyperref}
\usepackage{graphicx}%Include figure files
\usepackage{epstopdf}%ÓÃÓÚ¼ÓÔØͼƬ
\usepackage{amsmath}
\usepackage{amsfonts}
\usepackage{amssymb}
\usepackage{color}%
\usepackage{dcolumn}% Align table columns on decimal point
\usepackage{slashed}
\usepackage{amssymb,ulem}
\providecommand{\U}[1]{\protect\rule{.1in}{.1in}}

\newcommand{\f}{\begin{equation}}
\newcommand{\ff}{\end{equation}}
\newcommand{\fa}{\begin{eqnarray}}
\newcommand{\ffa}{\end{eqnarray}}

\begin{document}
%\begin{CJK*}{GBK}{kai}
\title{Chaotic dynamics of string around charged black brane with hyperscaling violation}
\author{Da-Zhu Ma $^{1}$}
\thanks{mdzhbmy@126.com}
\author{Dan Zhang $^{2}$}
\thanks{danzhanglnk@163.com}
\author{Guoyang Fu $^{3}$}
\thanks{FuguoyangEDU@163.com}
\author{Jian-Pin Wu $^{3,4}$}
\thanks{jianpinwu@yzu.edu.cn}
\affiliation{ $^1$ School of Information and Engineering, Hubei Minzu University, Enshi, 445000, China\ \\
$^2$ Department of Physics, School of Mathematics and Physics, Bohai University, Jinzhou 121013, China\ \\
$^3$ Center for Gravitation and Cosmology, College of Physical Science and Technology, Yangzhou University, Yangzhou 225009, China\ \\
$^4$ School of Aeronautics and Astronautics, Shanghai Jiao Tong University, Shanghai 200240, China}

\begin{abstract}
\vspace*{0.6cm}
\baselineskip=0.6 cm
By fast Lyapunov indicator (FLI), we study the chaotic dynamics of closed string around charged black brane with hyperscaling violation (HV).
The Hawking temperature, Lifshitz dynamical exponent and HV exponent together affect the chaotic dynamics of this system.
The temperature plays the role of driving the closed string to escape to infinity.
There is a threshold value $z_{\ast}=2$, below which the string is captured by the black brane no matter where the string is placed at the beginning.
However, when $z>2$, the string escapes to infinity if it is placed near the black brane at the beginning,
but if the initial position of string is far away from the black brane, it oscillates around the black brane till eternity,
which is a quasi-periodic motion.
HV exponent plays the role of driving the string falling into the black brane.
With the increase of HV exponent $\theta$, the falling velocity becomes faster.
We find that when we heat the system with large HV exponent, the chaotic system does not essentially changes.
It indicates that the HV exponent plays a very important role in determining the state of the chaotic system.
Also we study the the effect from the winding number of the string. The study indicates that the chaotic dynamics of the string is insensitive to the winding number.

\end{abstract} \maketitle

\section{Introduction}

The chaotic dynamics around black hole is one of the central attention in physics.
It exhibits many interesting phenomena and some important insight both at the astrophysical and the quantum level.
The simplest dynamics is the point particle, which is integrable in the generic Kerr-Newman background ~\cite{Carter:1968}.
In some complicated multi black hole geometries such as the Majumdar-Papapetrou geometry ~\cite{Majumdar:1947,Papapetrou:1947},
the motion of the point particle becomes chaotic ~\cite{Dettmann:1994dj,Hanan:2006uf}.
And then, lots of chaotic phenomena have also been discovered for the motions of charged particles in a magnetic field
interacting with gravitational waves ~\cite{Varvoglis:1992}, or particles near a black hole in a Melvin magnetic
universe ~\cite{Karas:1992}, or in a perturbed Schwarzschild spacetime ~\cite{Bombelli:1992,Aguirregabiria:1996vq,Sota:1995ms},
or in the accelerating and rotating black holes spacetime ~\cite{Chen:2016tmr}.

Different from the motion of a point particle, the motion of a string due to its extended nature,
the dynamics exhibits a more complex behavior.
Even on these radially symmetric spacetimes, the motion of the string is also chaotic ~\cite{Frolov:1999pj}.
More non-integrable behavior of string can be found in AdS spacetimes, see for example, ~\cite{Giataganas:2014hma,Giataganas:2013dha,Giataganas:2017guj}.
One of the motivations to study string dynamics arises from the AdS/CFT (Anti-de Sitter/Conformal Field theory) correspondence ~\cite{Maldacena:1997re,Gubser:1998bc,Witten:1998qj,Aharony:1999ti}.
In ~\cite{Zayas:2010fs}, the authors discover the chaotic behavior of the closed string around AdS-Schwarzschild (AdS-SS) black hole
by studying the power spectrum, the largest Lyapunov exponent, Poincare sections and basins of attractions.
And then, they propose that the operators being described by the closed string are generalizations of the Gubser-Klebanov-Polyakov ~\cite{Gubser:2002tv} operators and in particular, a positive largest Lyapunov exponent on the gravity side
corresponds to the appropriate bound for the time scale of Poincare recurrences on the gauge theory side.

Along this direction, the chaotic dynamics of the string have also been studied in more general geometries
in ~\cite{Ma:2014aha,Bai:2014wpa,Basu:2016zkr,Ishii:2016rlk,Cubrovic:2019qee}.
In the dynamical system of string around neutral Gauss-Bonnet (GB) black hole,
a critical value of GB parameter are found, below which the behavior of the system is
no-chaotic and above which it gradually becomes chaotic.
It is different from the case in the AdS-SS black hole, which is weakly chaotic ~\cite{Zayas:2010fs}.
The dynamics of string around neutral Lifshitz black hole has also been explored in ~\cite{Bai:2014wpa}
and they find that with the increase of the Lifshitz dynamical exponent,
the dynamical system becomes more chaotic.
Further, the authors in ~\cite{Basu:2016zkr,Cubrovic:2019qee} also study the chaotic behavior of the string around charged black hole.
They find that the temperature plays a crucial role in the generator of chaos.
With the increase of the temperature, the string system becomes more chaotic ~\cite{Basu:2016zkr,Cubrovic:2019qee}.

In this paper, we shall study the chaotic dynamics of the string around charged AdS black brane with hyperscaling violation (HV).
The HV spacetime is characterized by the Lifshitz dynamical critical exponent $z$ and HV exponent $\theta$ ~\cite{Fisher}.
Such scaling properties have been observed at criticality in many condensed matter system.
Lots of holographic systems with HV have also been constructed, see for example ~\cite{Gouteraux:2011ce,Huijse:2011ef,Dong:2012se,Charmousis:2010zz}.
In the HV theories, the entropy scales as $T^{(d-\theta)/z}$.
In contrast the theories with gravity dual having the standard AdS metric,
for which the entropy scales as $T^{d}$, we conclude that the HV results in an effective dimension $d_{\theta}=d-\theta$.
For a specific value $d-\theta=1$, the entanglement entropy springs up a logarithmic violation ~\cite{Wolf:2006zzb,Swingle:2009bf},
and results in an infrared metric which holographically describes a compressible state with hidden Fermi surfaces ~\cite{Huijse:2011ef}.
Further, many properties of the holographic dual system with HV, including the transports, (non-)Fermi behavior, etc.,
have been widely explored, see for example, ~\cite{Kuang:2014pna,Kuang:2014yya}.

Our work is organized as what follows. In Section ~\ref{section-b}, we present a brief introduction on the charged HV black brane.
In Section ~\ref{Section-string}, we derive the equations of motion of string in phase space.
And then, we investigate the properties of the string dynamics by numerically working out chaos indicators.
Finally in Section ~\ref{sec-con} we present our conclusions and some remarks on the future investigation.

\section {Charged black branes with hyperscaling violation}\label{section-b}

Our starting point is the following Einstein-Maxwell-Dilaton action in $4$ dimensional spacetimes ~\cite{Alishahiha:2012qu}
\fa
\label{action-HV}
S=-\frac{1}{16\pi G}\int d^{d+2}x\sqrt{-g}\Big[R-\frac{1}{2}(\partial\psi)^2+V_0 e^{\eta\psi}-\frac{1}{4}\Big(e^{\lambda_1\psi}F^2+e^{\lambda_2\psi}\mathcal{F}^2\Big)\Big]\,.
\ffa
The action contains two $U(1)$ gauge fields: $A$ with field strength $F_{\mu\nu}$ and $\mathcal{A}$ with $\mathcal{F}_{\mu\nu}$.
They both couple to a dilaton field $\psi$.
The gauge field $\mathcal{A}$ plays the role of an auxiliary field, which is necessary to generate an asymptotic Lifshitz geometry.
While the gauge field $A$ is the exact Maxwell field, which supports a charged black brane solution.
$\lambda_1$, $\lambda_2$, $\eta$ and $V_0$ are free parameters of the theory to be determined.

A charged black brane solution with hyperscaling violation from the above action can be given by ~\cite{Alishahiha:2012qu}
\begin{eqnarray}
&&
\label{Metric}
ds^{2}
=r^{-\theta}
\left[-r^{2z}f(r)dt^2+\frac{dr^2}{r^2 f(r)}
+r^2(dx^2+dy^2)\right]\,,
\
\\
&&
\label{fr}
f(r)=1-\frac{M}{r^{z+2-\theta}}+\frac{Q^2}{r^{2(z+1-\theta)}}\,,
\
\\
&&
\label{At}
A_t=\mu r_h^{-(z-\theta)}\Big(1-\Big(\frac{r_h}{r}\Big)^{z-\theta}\Big)\,,
\
\\
&&
\label{cAt}
\mathcal{A}_t=-\slashed{\mu}r_h^{2+z-\theta}\Big(1-\Big(\frac{r}{r_h}\Big)^{2+z-\theta}\Big)\,.
\end{eqnarray}
$r_h$ is the radius of horizon satisfying $f(r_h)=0$.
$M$ and $Q$ are the mass and charge of the black brane, respectively.
There is a relation between them
\fa
r_h^{2(z-\theta+1)}-M r_h^{z-\theta}+Q^2=0\,,
\ffa
which is given by $f(r_h)=0$.
$\mu$ and $\slashed{\mu}$ are defined as
\fa
&&
\label{mu}
\mu=Q\sqrt{\frac{2(2-\theta)}{z-\theta}}\,,
\
\\
&&
\label{mu-s}
\slashed{\mu}=\sqrt{\frac{2(z-1)}{2+z-\theta}}\,.
\ffa
In the holographic framework, $\mu$ denotes the chemical potential in the boundary field theory.
All the parameters of the theory depend on the Lifshitz scaling exponent $z$ and HV
exponents $\theta$ and they can be written as
\fa
\label{parameters}
&&
\lambda_1=\sqrt{\frac{2(z-1-\theta/2)}{2-\theta}}\,,
\nonumber
\\
&&
\lambda_2=-\frac{2(2-\theta/2)}{\sqrt{2(2-\theta)(z-\theta/2-1)}}\,,
\nonumber
\\
&&
\eta=\frac{\theta}{\sqrt{2(2-\theta)(z-1-\theta/2)}}\,,
\nonumber
\\
&&
V_0=(z-\theta+1)(z-\theta+2)\,.
\ffa
The Hawking temperature of the black brane is
\fa
\label{HT-v1}
\hat{T}=\frac{(2+z-\theta)r_h^z}{4\pi}\Big[1-\frac{(z-\theta)Q^2}{2+z-\theta}r_h^{2(\theta-z-1)}\Big]\,.
\ffa
This black brane solution has the following scaling symmetry:
\begin{eqnarray}\label{rescaling}
&& r\rightarrow r_h r~,~~~t\rightarrow
\frac{t}{r_h^z}~,~~~(x,y)\rightarrow \frac{1}{r_h}(x,y)~,~~~
T\rightarrow
\frac{T}{r_h^z}~,
\nonumber\\
&& Q\rightarrow r_h^{(z-\theta+1)}Q~,~~~A_t\rightarrow r_h
A_t~,~~~\mathcal{A}_t\rightarrow r_h^{\theta-z-2} \mathcal{A}_t~,
\end{eqnarray}
under which, we can set $r_h=1$.
Now, we can reexpress the redshif factor, the gauge fields and the Hawking temperature as
\fa
&&
\label{frv1}
f(r)=1-\frac{1+Q^2}{r^{z+2-\theta}}+\frac{Q^2}{r^{2(z+1-\theta)}}\,,
\
\\
&&
\label{Atv1}
A_t=\mu \Big[1-\Big(\frac{1}{r}\Big)^{z-\theta}\Big]\,,
\
\\
&&
\label{cAtv1}
\mathcal{A}_t=-\slashed{\mu}\Big(1-r^{2+z-\theta}\Big)\,,
\
\\
&&
\label{HTv2}
\hat{T}=\frac{(2+z-\theta)}{4\pi}\Big[1-\frac{(z-\theta)Q^2}{2+z-\theta}\Big]\,.
\ffa
For given model parameters $z$ and $\theta$, this black brane is specified by one scaling-invariant parameter
$\hat{T}/\mu$, for which we abbreviate it to $T$.

Before proceeding, it is important to fix the allowed region of $z$ and $\theta$.
First, the black brane solution \eqref{Metric}, \eqref{fr}, \eqref{At} and \eqref{cAt} are
valid only for $z\geq 1$ and $\theta\geq 0$. The
geometry with $z=1$ and $\theta=0$ reduces to the
AdS one. Second, to have a well-defined chemical potential in the dual field theory,
the requirement $z-\theta\geq0$ shall be imposed.
Third, the null energy condition impose that
$(-\frac{\theta}{2}+1)(-\frac{\theta}{2}+z-1)\geq0$ ~\cite{Alishahiha:201209}.
Combining the condition $\theta< 2$, which results from Eq. \eqref{mu}, we have $\theta\leq2(z-1)$.
Based on the above analysis, the allowed range of the model parameters is given by
\begin{eqnarray}\label{ParameterRegion}
\left\{
\begin{array}{rl}
&0\leq \theta \leq 2(z-1) \quad {\rm for} ~~~1\leq z<2   \ ,   \\
&0\leq\theta<2 \quad {\rm for}~~~ z\geq2  \ .
\end{array}\right.
\,
\end{eqnarray}

\section{Closed string around charged HV black branes}\label{Section-string}

A closed string can be described by the following Polyakov action,
\begin{eqnarray}\label{Polyakovaction}
\mathcal{L}=-\frac{1}{2\pi \beta}\sqrt{-g}g^{\mu\nu}G_{ab}\partial_\mu X^a\partial_\nu X^b,
\end{eqnarray}
where $\beta$ is the coupling, which relates the string length $l_s$ by $l_s^2=\beta$.
$X^a$ is the coordinates of the target space with the metric $G_{ab}$.
We parameterize the worldsheet of the string by the coordinates $\sigma^\mu=(\tau, \sigma)$.
$g_{\mu\nu}$ is the induce metric on the worldsheet. We can work in the conformal gauge $g_{\mu\nu}=\eta_{\mu\nu}$.
But at this moment, the Virasoro constraints must be supplemented,
%%%%%%%
\fa
\label{virasoro-constraint}
G_{ab}\left(\partial_\tau{X}^a\partial_\tau{X}^b+\partial_\sigma X^{a}\partial_\sigma X^{b}\right)=0\,,~~~~~~G_{ab}\partial_\tau{X}^a\partial_\sigma X^{b}=0\,.
\ffa
%%%%%%%
The first constraint is just the Hamiltonian constraint $H=0$.
For closed string, $0\leq\sigma\leq 2\pi$.

In this paper, we only focus on the dynamics of the closed string in charged black branes with HV in $4$ dimension.
It is convenient to study the dynamics of the closed string in polar coordinates, in which the horizon manifold is wrote as
\f
\label{hman}
d\vec{x}^2_{2}=d\rho^2+\rho^2 d\phi^2\,.
\ff
We consider the following consistent ansatz:
\begin{eqnarray}\label{Embedding_AdS_4}
t=t(\tau),~~r=r(\tau),~~\rho=\rho(\tau),~~\phi=\alpha\sigma\,.
\end{eqnarray}
$\alpha$ is the winding number, which describes the differences between strings and particles.
The integrability and chaos of strings with the above form have been well studied in ~\cite{Basu:2011fw,Frolov:1999pj,Zayas:2010fs,Basu:2011di,Basu:2011dg,Basu:2016zkr,Ma:2014aha,Cubrovic:2019qee}.
In this setup, we can explicitly write the Polyakov Lagrangian as
\begin{eqnarray}\label{PolyakovactionConcrete_AdS_4}
\mathcal{L}=-\frac{r^{-\theta}}{2\pi \beta}\left[r^{2z} f(r)\dot{t}(\tau)^2-\frac{\dot{r}(\tau)^2}{r^2 f(r)}+r^2\left(-\dot{\rho}(\tau)^2+\rho^2\alpha^2 \right)\right]\,,
\end{eqnarray}
where the dot denotes the derivative with respect $\tau$ (the same hereinafter)
and we will represent the derivative with respect to $r$ by using prime in what follows.
Using the canonical momentum transform, we can give the Hamiltonian as
\begin{eqnarray}\label{Hamiltonian_AdS_4}
H = \frac{\pi\beta}{2}r^{\theta} \bigg[r^2 f(r) p_r^2 - \frac{1}{r^{2z} f(r)} p_t^2 +\frac{p_{\rho}^2 }{r^2}\bigg]+\frac{1}{2 \pi \beta} r^{2-\theta} \alpha^2 \rho^2\,.
\end{eqnarray}
The Hamiltonian satisfies the constraint $H=0$, which is just the first equation in Eq.\eqref{virasoro-constraint}.
$p_t$, $p_r$ and $p_{\rho}$ are the canonical momentums, which are
\begin{eqnarray}\label{CMomentum_AdS_4}
p_t= - \frac{r^{2 z-\theta} f(r) \dot{t}(\tau)}{\pi \beta}, \quad p_r= -\frac{r^{-(2+\theta)} \dot{r}(\tau)}{\pi \beta f(r)}, \quad
p_{\rho}=\frac{r^{2-\theta} \dot{\rho}(\tau)}{\pi \beta} .
\end{eqnarray}
They construct the canonical phase space $\{t,p_t\}$, $\{r,p_r\}$ and $\{\rho,p_{\rho}\}$.
Then, using the Poisson bracket, we derive the canonical equations of motion, which are
\fa
&&
\label{tEOM_AdS_4}
\dot{t}:=\{t,H\}=-\frac {\pi  r^{\theta - 2 z} \beta} {f (r)}  p_t\,,
\\
&&
\label{ptEOM_AdS_4}
\dot{p}_t:=\{p_t,H\}=0\,,
\\
&&
\label{rEOM_AdS_4}
\dot{r}:=\{r,H\}=\pi \beta f (r) r^{2+ \theta} p_r\,,
\\
&&
\label{prEOM_AdS_4}
\dot{p_r}:=\{p_r,H\}\,,
\nonumber
\\
&&
=
\frac{r^{1-\theta} \alpha^2 (\theta -2)}{2 \pi \beta}\rho^2-\left(\frac{\pi  r^{\theta-2z-1} \beta (2z - \theta)}{2 f(r)}+ \frac{\pi r^{\theta- 2z } \beta f'(r)}{2 f(r)^2}\right)p_t^2
-\frac{1}{2} \pi r^{ \theta-3} \beta (\theta -2) p_\rho^2
\nonumber
\\
&&
- \frac{r^{1+\theta}}{2} \pi \beta\left((2+\theta)f(r)+r f'(r)\right) p_r^2\,,
\\
&&
\label{rhoEOM_AdS_4}
\dot{\rho}:=\{\rho,H\}=\pi \beta  r^{\theta - 2} p_ {\rho}\,,
\\
&&
\label{prhoEOM_AdS_4}
\dot{p_\rho}:=\{p_\rho,H\}=- \frac{r^{2-\theta} \alpha^2 }{\pi \beta} \rho\,.
\ffa
From Eq.\eqref{ptEOM_AdS_4}, we have a constant of motion $p_t=E$,
which relates to the energy.

\section{Chaotic dynamics of string in HV charged black brane}\label{sec-dynamics}

\subsection{Introduction of numerical method}

A reliable numerical technique is very essential to solve the EOMs of a closed string.
In this subsection, we firstly present a brief introduction on the numerical method we adopted in what follows.

Over the past few decades, a set of new numerical techniques have been developed to solve ordinary differential equations,
such as manifold corrections and symplectic algorithms.
The application of symplectic algorithm is limited because it requires the Hamiltonian can be separated into two individual parts.
Besides, for high-order symplectic algorithms, the canonical difference scheme should be given beforehand.
Based on this, we only focus on the manifold correction method in this paper.

In general, the classical forth-order Runge-Kutta algorithm(RK4) is used to solve the Hamiltonian Eq.\eqref{Hamiltonian_AdS_4}.
But abundant numerical simulations indicate that the error of $\Delta H$ obtained from RK4 is nearly the linear growth with time,
that is, the integrated coordinates and momentums have deviated the original hyper-surface.
In order to pull the deviated hyper-surface back,
one of the rigorous manifold correction methods called the velocity scaling method ~\cite{NewAMa} is adopted.
The accuracy of the solution for this method reported in the references ~\cite{Ma:2014aha,IJMPCMa,APJMa} is well
because it puts the deviated hypersurface back with a least-squares shortest path.
In this paper, the Hamiltonian of the system could be treated as a conserved quantity because it is subjected to the constraint {$H=0$}.
We assume that there is a relation between the numerical solutions $p_r$ and $p_{\rho}$ from RK4 and the corrected values $p_r^{\ast}$ and $p_{\rho}^{\ast}$ as what follow
\begin{eqnarray}\label{manifold_AdS_4}
p_r^{\ast}=\gamma p_r,  \quad p_{\rho}^{\ast}=\gamma p_{\rho}\,,
\end{eqnarray}
where $\gamma$ is a dimensionless parameter.
Substituting Eq.\eqref{manifold_AdS_4} into Eq.\eqref{Hamiltonian_AdS_4},
the form of $\gamma$ is
\begin{eqnarray}\label{rr_AdS_4}
\gamma=\sqrt{\frac{\Big[\frac{2\big(H-\frac{r^{2-\theta}\alpha^{2}\rho^{2}}{2\pi\beta}\big)}{\pi\beta r^{\theta}}+\frac{E^{2}}{r^{2z}f(r)}\Big]}{r^{2}f(r)p_{r}^{2}+\frac{p_{\rho}^{2}}{r^{2}}}}\,.
\end{eqnarray}
After this treatment, the error of the Hamiltonian can be controlled well.
The error is almost kept in double precision of the machine $10^{-16}$ at every integration step.
Thus, it provides sufficient accuracy in computations.

The evolution of the chaotic system is sensitive to its initial conditions.
Based on this remarkable characteristic, a great number of chaos indicators have been developed.
They are bifurcations ~\cite{Buchler1987,Maciejewski2001}, fractal basin boundaries ~\cite{Levin2000}, Poincar\'{e} sections ~\cite{Markov1980},
Lyapunov exponent (LE) ~\cite{Eckmann1985,Edward1993,Huang2014,Skokos2010},
fast Lyapunov indicator (FLI) ~\cite{Froeschle1997,Froeschle2000,Wu2003,Huang2014},
relative finite-time Lyapunov indicator (RLI) ~\cite{Sandor2000,Sandor2004},
smaller alignment index (SALI) ~\cite{Skokos2001,Skokos2004,Soulis2007,Soulis2008,Bountis2009},
and generalized alignment index (GALI) method ~\cite{Skokos2007,Skokos2016}, etc.
About these chaos indicators, each of them has its advantages and disadvantages.
The method of bifurcation is easy to find a period doubling, quadrupling, etc., and the onset of chaos.
The method of fractal basin boundaries is unsuitable for the prediction of instability (or chaos) of orbits in compact binaries systems ~\cite{Wu2015}.
Poincar\'{e} sections method is appropriate for not more than four-dimensional systems.
In general, LE, FLI, RLI, SALI, and GALI are useful indicators in conservative systems.
They have been widely used in n-body problems, cosmology and general relativity models ~\cite{Ma2016}.
Especially, LE is also effective in dissipative system, but FLI, RLI, SALI, and GALI are not ~\cite{Ma2016}.
LE estimates order or chaos by means of measuring the separation of two nearby orbits.
Using the variational method or the two-particle method, the value of LE is easy to be obtained.
It should be noted that only the invariant LE with independent coordinate is applicable for
the Mixmaster cosmology and the conservative second Post-Newtonian Lagrangian dynamics of spinning compact binaries ~\cite{Wu2015}.
RLI is faster than LE, and it is suitable for symplectic mappings or continuous Hamiltonian systems.
LE and RLI, they need only one deviation vector with respect to the reference orbit.
However, two deviation vectors are required for SALI.
GALI, as an improved version of SALI, it is faster and reveals much more useful information than SALI on the local dynamics.
Compared with LE, RLI, SALI, and GALI, FLI is the fastest because it does not need to compute a limit value.
What's more, FLI does not have to use the variation equation.
Considering the calculation of the variational equations is complicated in this paper,
we only pay attention to FLI.

As a variant of the maximum LE, and without the renormalization and the average versus time, FLI is quicker and easier to use.
The form is ~\cite{Wu2003} ,
\begin{eqnarray}\label{FLI}
\textbf{FLI}(t)=\log_{10}\frac{\|\textbf{d}(t)\|}{\|\textbf{d}(0)\|}.
\end{eqnarray}
Where $\textbf{d}(0)$ and $\textbf{d}(t)$ represent the distances at the starting point and time t,
$d(0)$ is about $10^{-9}$.
It is obviously to see that FLI curves of the bounded orbit display different behaviors for chaotic and regular motion,
an exponential rate of divergence of two nearby trajectories represents the orbit is chaotic,
while a polynomial divergence reveals the orbit is regular.

\subsection{Chaotic dynamics}\label{app-dynamics}

In this section, we shall numerically study the chaotic dynamics of closed string.

\subsubsection{The role of the temperature}

To clearly study what role the temperature plays, we set $\alpha=0$, $z=1$ and $\theta=0$ in this section.
When $z=1$ and $\theta=0$, the HV black brane reduces to the RN black brane, around which the chaotic dynamics has been studied in ~\cite{Basu:2016zkr}.
In ~\cite{Basu:2016zkr}, they study the phase space $(r, p_{r})$ and the time evolution of the string.
Also, they study the chaotic indicators, including fractal basin boundaries and LLE.
But here, for self-consistency of our paper and clarifying some problems, we also exhibits the dynamics of closed string around RN-AdS black brane.

\begin{figure}
	\includegraphics[scale=0.65, bb= 0 0 265 200]{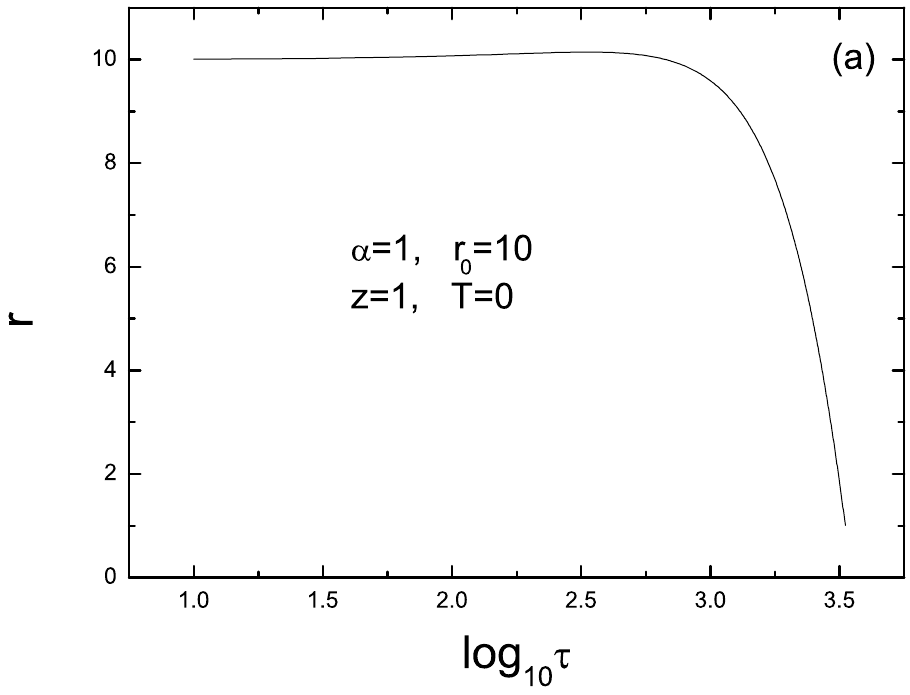}\ \hspace{0.8cm}
    \includegraphics[scale=0.65, bb= 0 0 270 195]{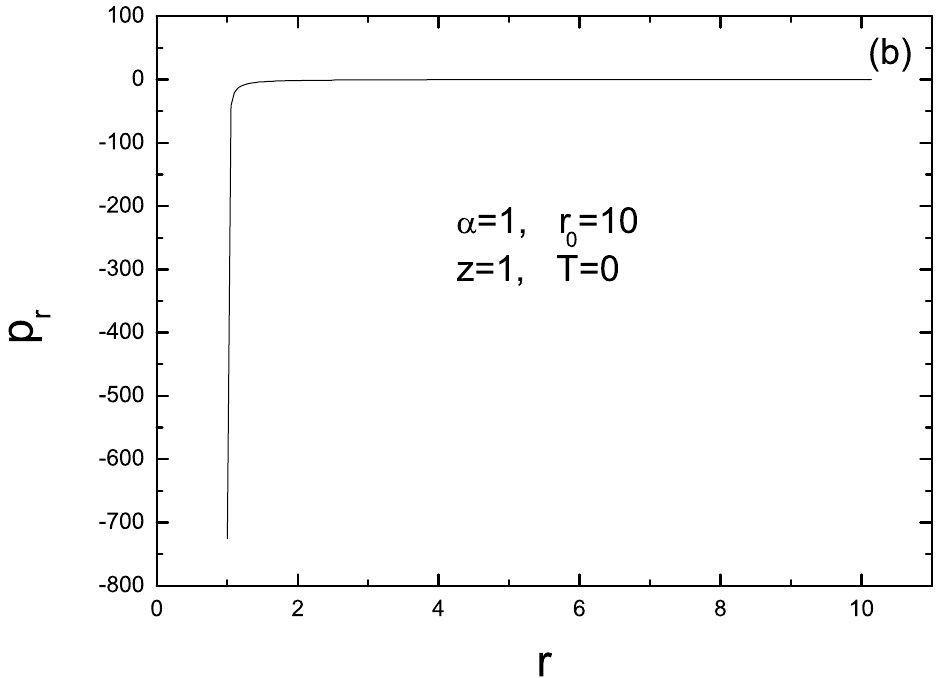}\ \\
    \includegraphics[scale=0.65, bb= 0 0 280 210]{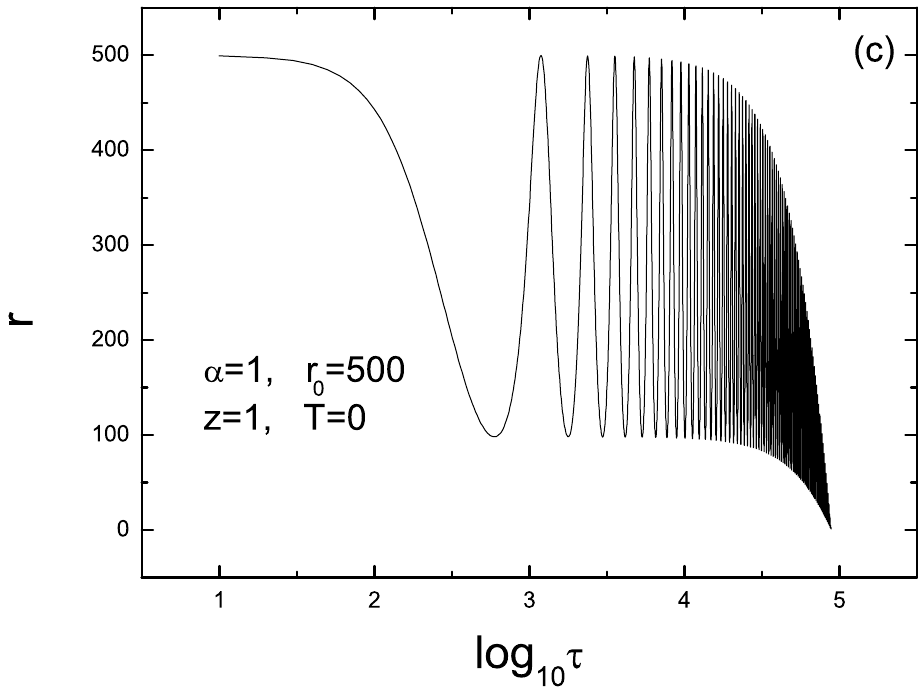}\ \hspace{0.8cm}
    \includegraphics[scale=0.65, bb= 0 0 280 210]{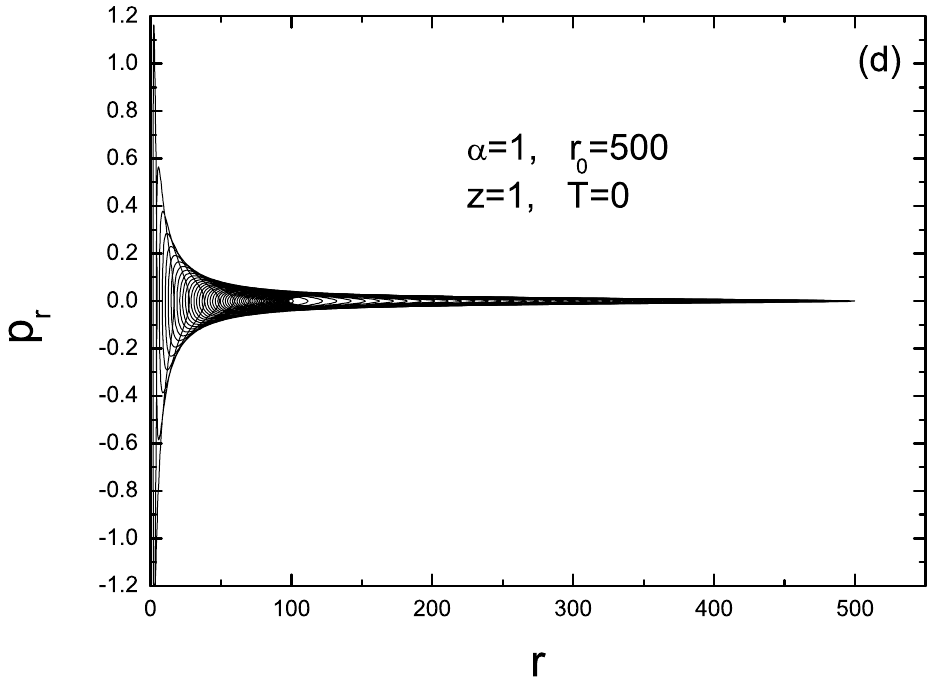}\ \\
    \caption{The time evolution of $r(\tau)$ (left panels) and the phase space $(r, p_{r})$ (right panels)
    with different initial conditions (the panels above are $r_0=10$ and that below are $r_0=500$) at $T=0$.
    Here, we have set $\alpha=1$, $\theta=0$, $z=1$, $E=5$, $\beta=1/\pi$, $\rho=0.01$ and $p_{\rho}=0$.
    }
    \label{fig:example_figure_v1}
\end{figure}
%%%%%%%%%%%
\begin{figure}
	\includegraphics[scale=0.65, bb= 0 0 280 210]{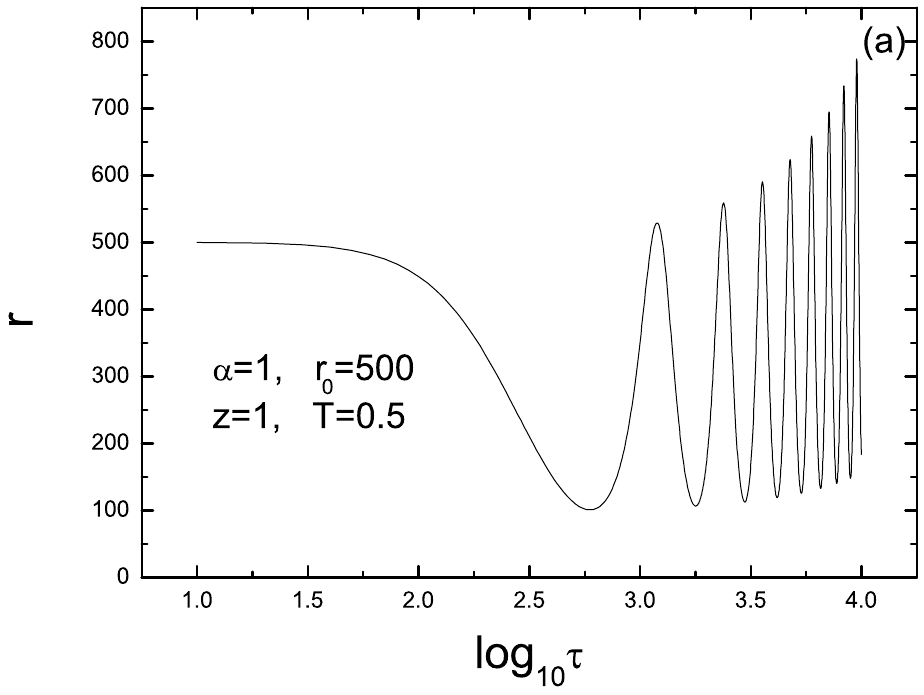}\ \hspace{0.8cm}
    \includegraphics[scale=0.65, bb= 0 0 280 210]{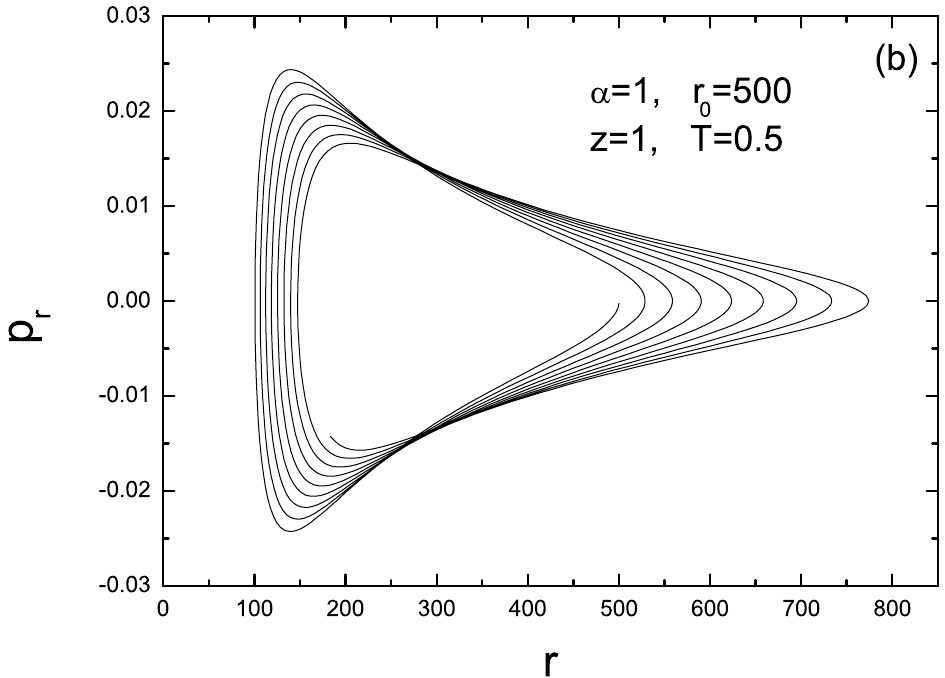}\ \\
    \includegraphics[scale=0.65, bb= 0 0 280 210]{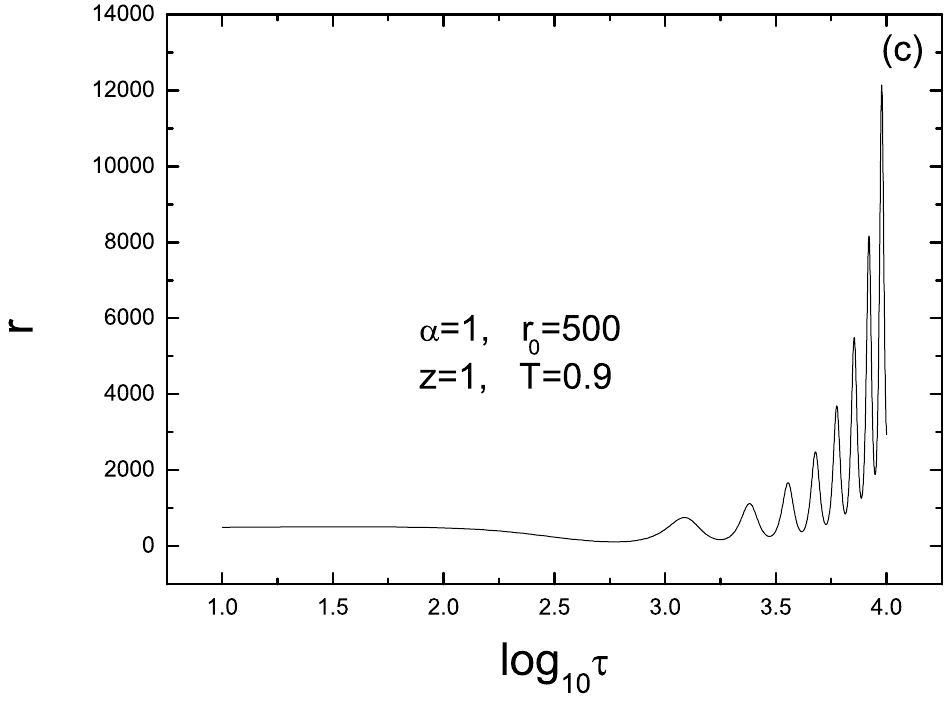}\ \hspace{0.8cm}
    \includegraphics[scale=0.65, bb= 0 0 280 210]{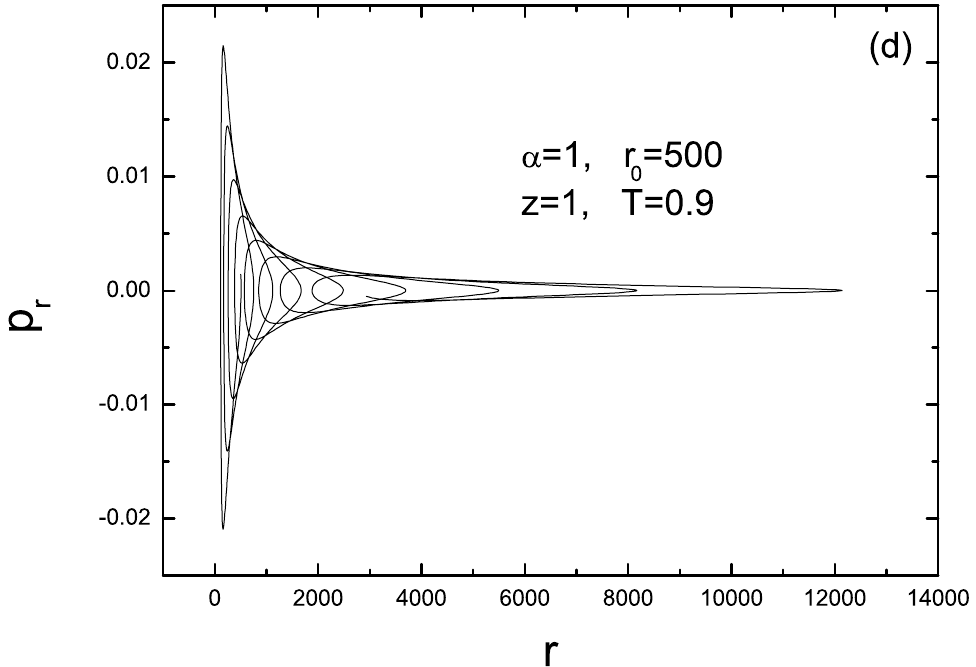}\ \\
    \caption{
    The time evolution of $r(\tau)$ (left panels) and the phase space $(r, p_{r})$ (right panels)
    with $r_{0} = 500$ for different temperatures (the panels above are $T=0.5$ and that below are $T=0.9$).
    Here, we have set $\alpha=1$, $\theta=0$, $z=1$, $E=5$, $\beta=1/\pi$, $\rho=0.01$ and $p_{\rho}=0$.
    }
    \label{fig:example_figure_2}
\end{figure}
%%%%%%%%%%%%%%%%

To gain a visual picture of dynamics of closed string, we firstly exhibit the time evolution of $r(\tau)$ and
the dynamics of string in the phase space $(r, p_{r})$.
Without loss of generality, we shall set $E=5$, $\rho=0.01$ and $p_{\rho}=0$ in our paper.

FIG.\ref{fig:example_figure_v1} shows the time evolution of $r(\tau)$ and the dynamics of closed string in the phase space $(r, p_{r})$ at $T=0$.
We find that when the initial position of closed string is close to the black brane (the plots above in FIG.\ref{fig:example_figure_v1}),
the string rapidly falls toward the black brane and is captured by the black brane finally.
When we place the closed string far away from the black brane,
the string firstly oscillates and after oscillating finite times, it also falls into the black brane.
As the initial position of closed string increases, the string oscillating time becomes longer.
When $r_0=500$, the oscillating time has climbed up to $10^{4}$ (the left below in FIG.\ref{fig:example_figure_v1}).
Therefore, we conclude that at zero temperature, regardless of the initial position the final destination of the string is falling into the black brane.

\begin{figure}
	\includegraphics[scale=0.65, bb= 0 0 280 210]{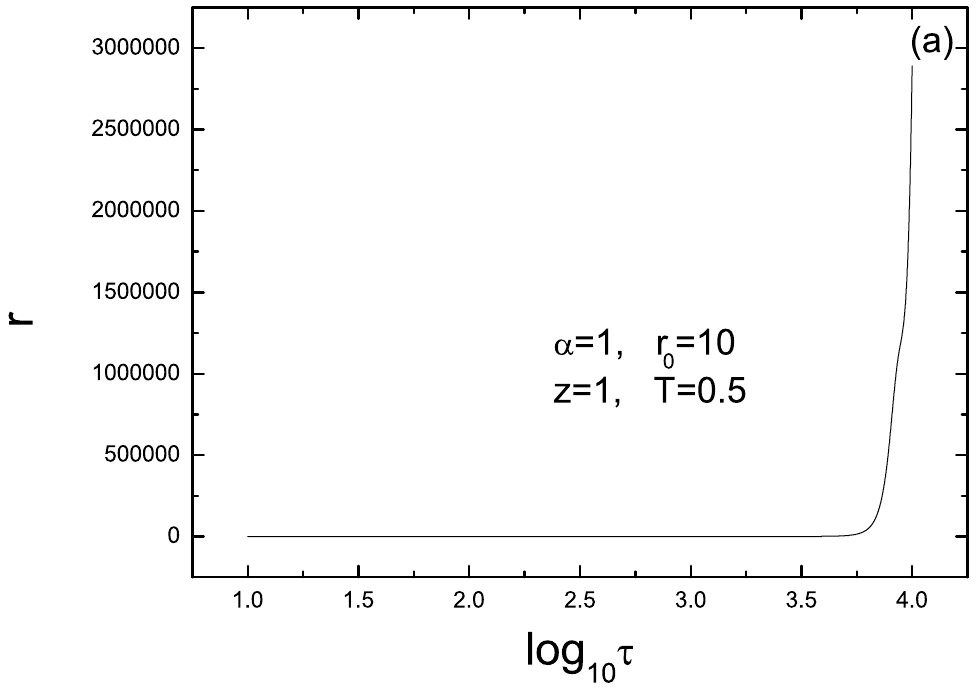}\ \hspace{0.8cm}
    \includegraphics[scale=0.65, bb= 0 0 280 210]{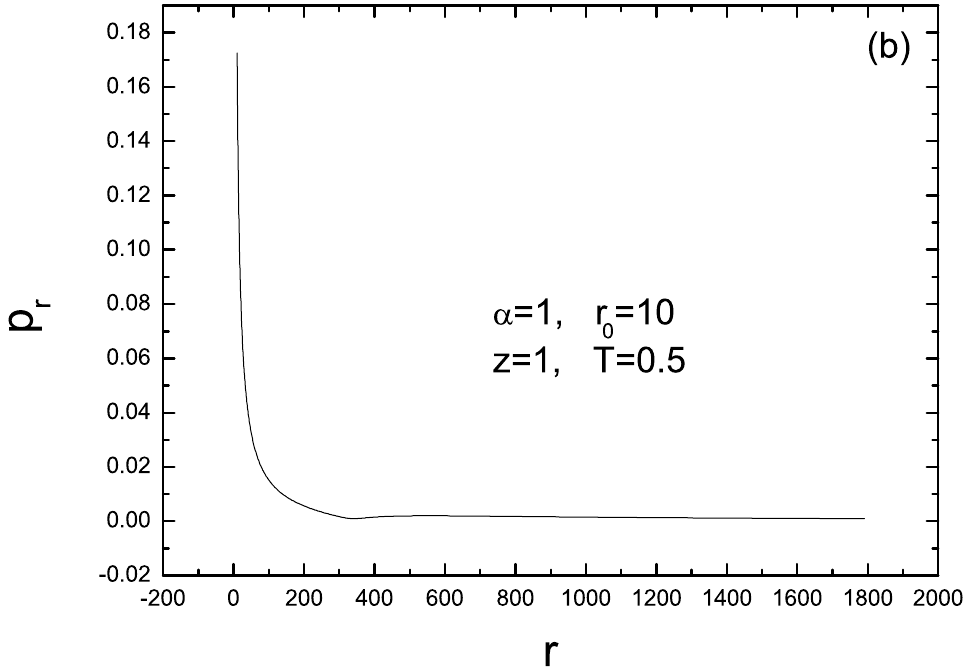}\ \\
    \includegraphics[scale=0.65, bb= 0 0 280 210]{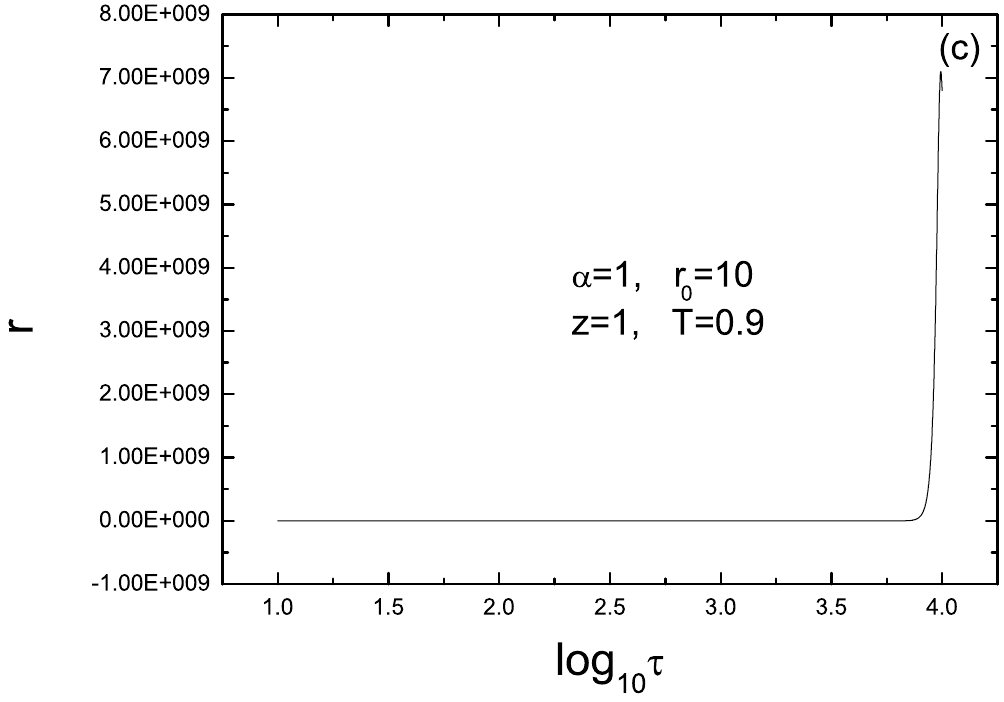}\ \hspace{0.8cm}
    \includegraphics[scale=0.65, bb= 0 0 280 210]{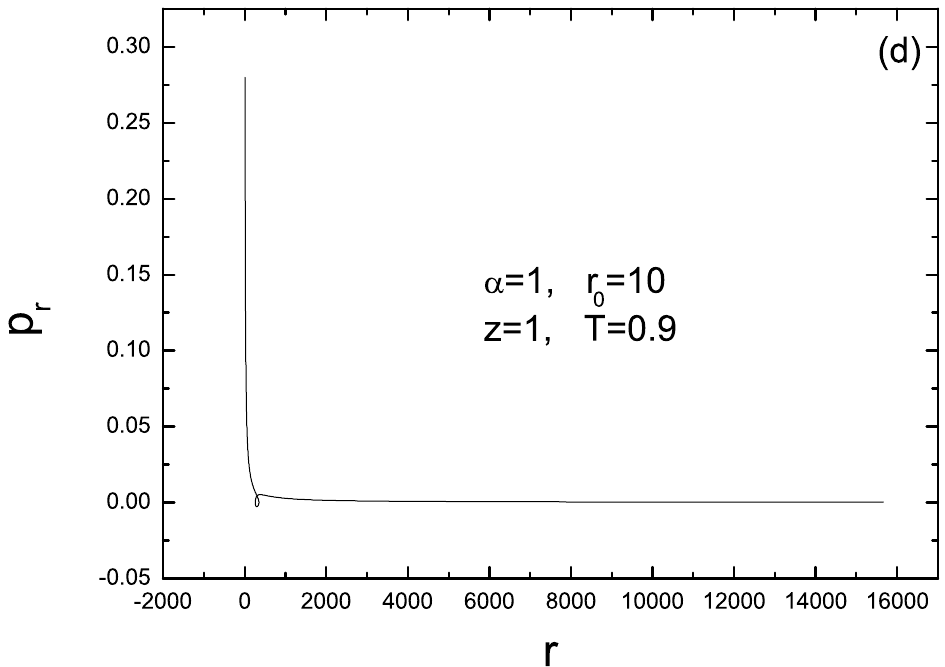}\ \\
    \caption{The time evolution of $r(\tau)$ (left panels) and the phase space $(r, p_{r})$ (right panels)
    with $r_{0} = 10$ for different temperatures (the panels above are $T=0.5$ and that below are $T=0.9$).
    Here, we have set $\alpha=1$, $\theta=0$, $z=1$, $E=5$, $\beta=1/\pi$, $\rho=0.01$ and $p_{\rho}=0$.}
    \label{fig:example_figure_3}
\end{figure}

Now, we heat the system.
We firstly discuss the case of large initial position of the closed string ($r_{0}=500$).
The panels above in FIG.\ref{fig:example_figure_2} show the dynamical evolution of closed string for $T=0.5$.
We see that the closed string oscillates at the beginning and then escapes to infinity slowly,
which is different from that at $T=0$.
As the temperature $T$ further is increased, the escape velocity of the closed string becomes faster
(see the panels below in FIG.\ref{fig:example_figure_2}, in which the temperature is $T=0.9$).
Further, we find that even when the closed string is at the position being close to the black brane (here, we take $r_0=10$),
the string also escapes to infinity rapidly as the temperature is further increased (see the FIG.\ref{fig:example_figure_3}).
Therefore, we conclude that the temperature plays the role of driving the closed string to escape to infinity. In particular, as the temperature increases, the escape velocity becomes faster.

Although the phase curve $(r, p_{r})$ and the time evolution of $r(\tau)$ can provide a visual picture of dynamics of closed string,
this method is complicated and cumbersome in the numerics.
In order to get a better perspective of the closed string dynamics,
we can explore the dynamics of the closed string by adopting some of chaos indicators.
In this paper, we study the chaotic dynamics by using FLIs£¬which is a high-efficient method in dealing with
the complicated calculation of the variational equations.

\begin{figure}
	\includegraphics[scale=0.7, bb= 0 0 280 210]{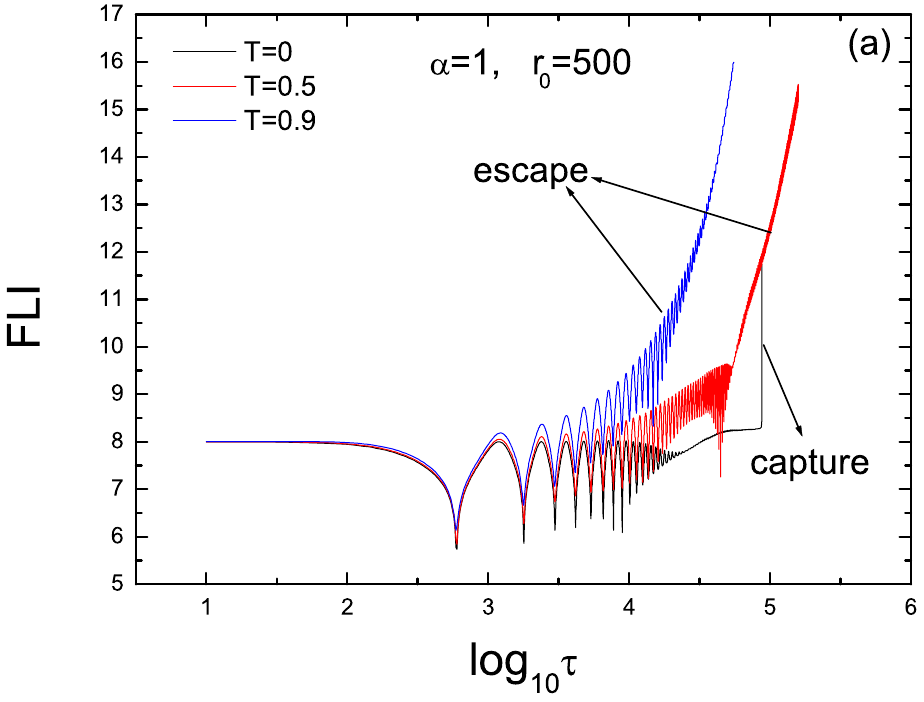}\ \hspace{0.8cm}
    \includegraphics[scale=0.7, bb= 0 0 280 210]{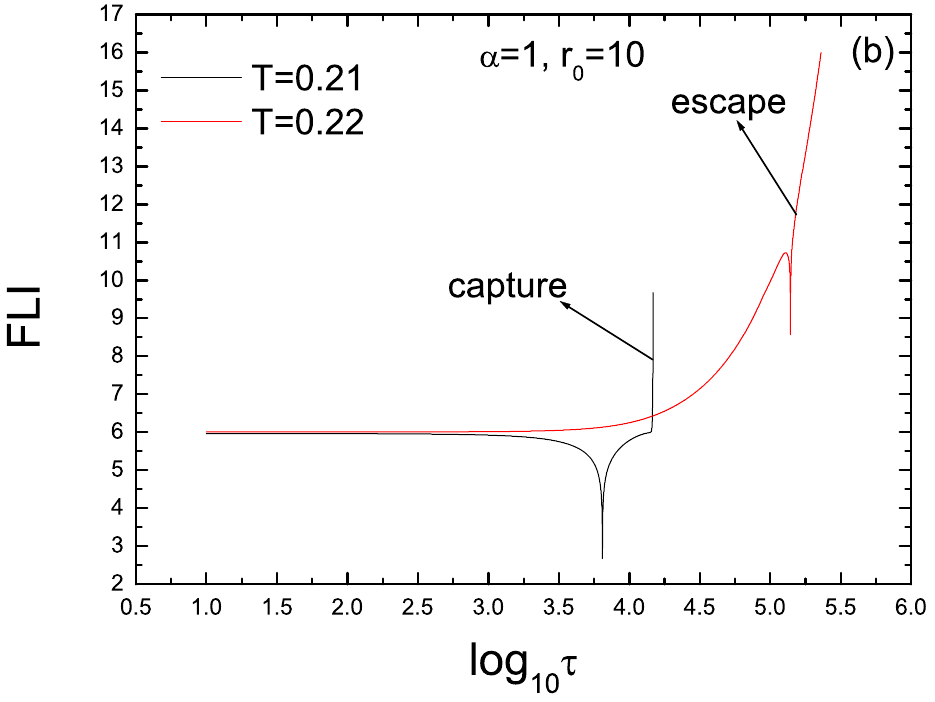}\ \\
    \caption{FLIs for $r_{0}=500$ (left plot) and $r_{0}=10$ (right plot) with different temperatures $T$. Here, $\alpha=1$, $z=1$, $\theta=0$.}
    \label{fig_FLIs_ze1}
\end{figure}

FIG.\ref{fig_FLIs_ze1} depicts the FLIs for $r_{0}=500$ (left plot) and $r_{0}=10$ (right plot) with different temperatures $T$.
When the string is placed far away from the black brane, we find that the curves of FLIs are almost the same at the initial phase of the evolution.
It implies that the ring string has a finite number of oscillations around the black hole.
However, in the later stage of the evolution, FLIs exhibits very different behavior for different temperatures.
For $T=0$, the curve is parallel with the vertical coordinate, which indicates that the closed string has been captured by the black brane.
For $T=0.5$ and $T=0.9$, both the two curves have an exponential growth with time,
which means that the closed string escapes to infinity in those two cases.
Especially, the deviation of the FLIs curves from the vertical coordinate becomes more significant for $T=0.9$ than $T=0.5$.
It again confirms that when we heat the system, the escaping velocity of the closed string becomes faster.
Also, we plot the FLIs for $r_0=10$ (right plot in FIG.\ref{fig_FLIs_ze1}),
which also clearly exhibits the two behaviors of closed string: escaping to infinity or being captured by the black brane depending on the temperature of the black brane.
Note that before escaping to infinity or being captured by the black brane, the string undergoes a longtime oscillation for $r_0=500$ but not for $r_0=10$.
On the contrary, the escaping or falling velocity is rapid for $r_0=10$. These observations are consistent with that in the time evolution and the phase space.
But we would like to point out that FLIs provide a more convenient and straightforward than the methods of the time evolution and the phase space
and we shall adopt FLIs to find more valuable consequences of this system.

\subsubsection{The effect from Lifshitz dynamical exponent}

In this subsection, we mainly study the effect from the Lifshitz dynamical exponent.
So we fix $\alpha=1$ and $\theta=0$.

\begin{figure}
	\includegraphics[scale=0.7, bb= 0 0 280 210]{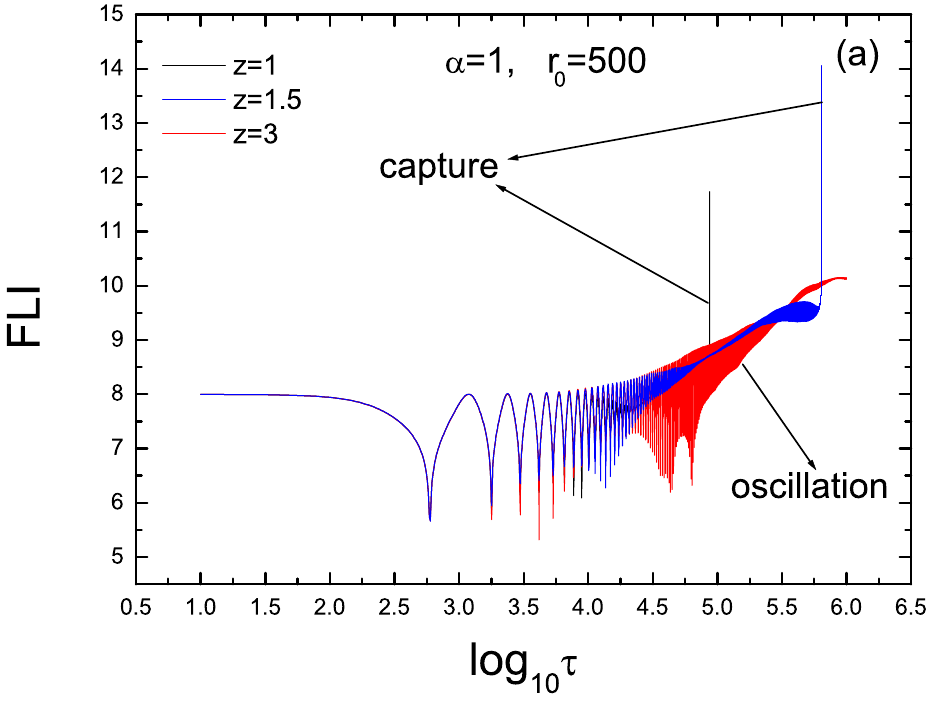}\ \hspace{0.8cm}
    \includegraphics[scale=0.7, bb= 0 0 280 210]{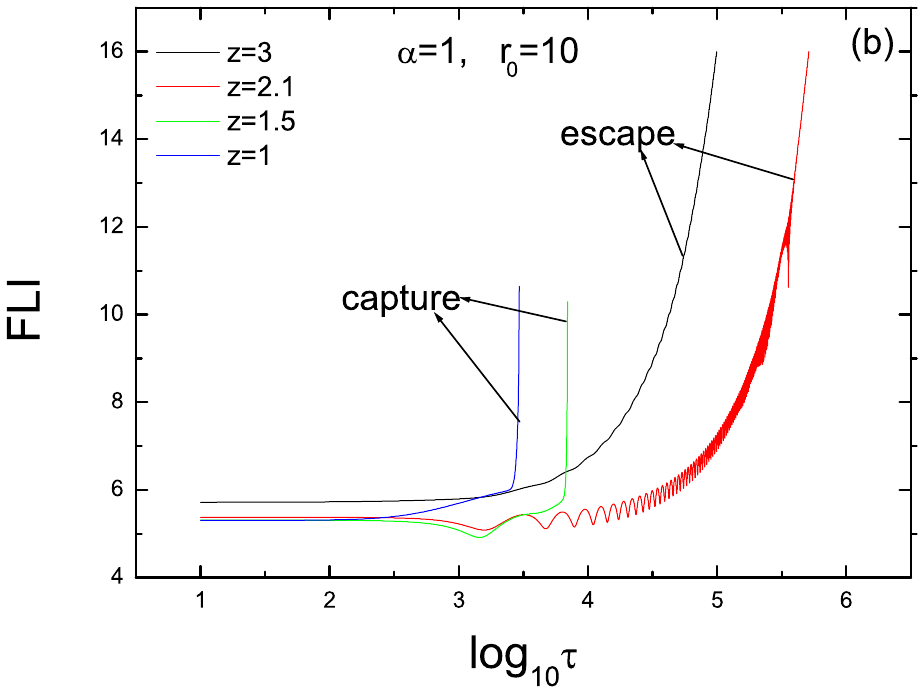}\ \\
    \caption{FLIs for $r_{0}=500$ (left plot) and $r_{0}=10$ (right plot) with different $z$. Here, we set $\alpha=1$ and $T=0$.}
    \label{fig_FLIs_dz}
\end{figure}

Firstly, we study the case of the extremal black brane, i.e., $T=0$.
The left plot in FIG.\ref{fig_FLIs_dz} shows FLIs with sample Lifshitz dynamical exponent $z$ for $r_{0}=500$ at zero temperature.
We find that the closed string is captured by the black brane after a finite number of oscillations for $z=1$ and $z=1.5$,
but oscillates around the black brane till eternity for $z=3$, which indicates that the motion is quasi-periodic.
Further, we change Lifshitz dynamical exponent $z$ from $1$ to $5$ with the same step $\Delta z =0.1$ to observe the FLIs
and we find that there is a threshold value $z_{\ast}=2$, under which the string is captured by the black brane after a finite number of oscillations
and beyond which it oscillates around the black brane till eternity.
When the initial position of the string is close to the black brane ($r_{0}=10$ here), we find that
no matter what $z$ changes, the quasi-periodic motion does not appear.
The string either is captured ($z<2$) or escapes to infinity ($z>2$) (see the right plot in FIG.\ref{fig_FLIs_dz}).

\begin{figure}
	\includegraphics[scale=0.7, bb= 0 0 280 210]{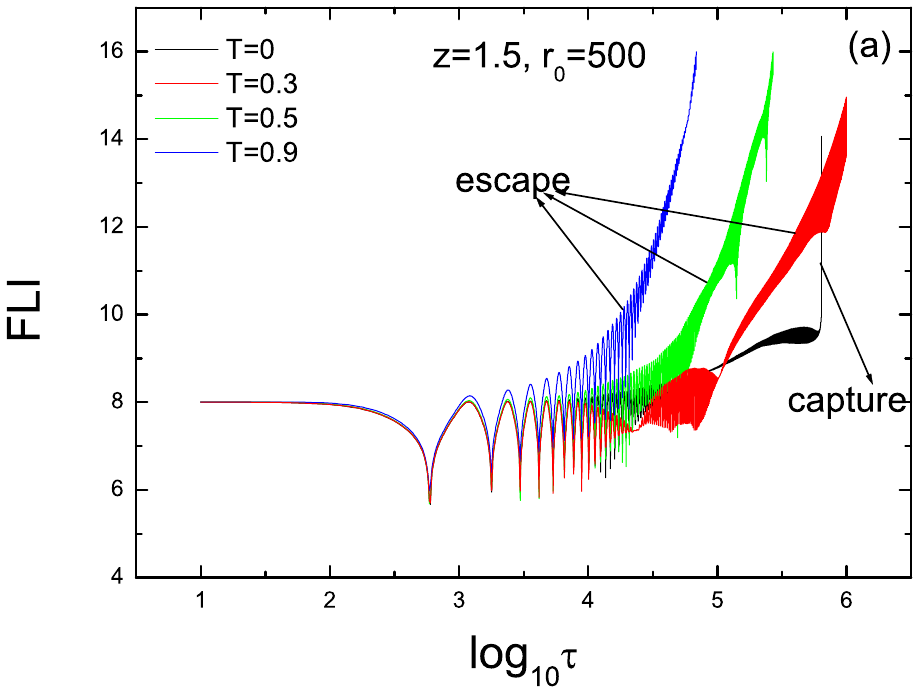}\ \hspace{0.8cm}
    \includegraphics[scale=0.7, bb= 0 0 280 210]{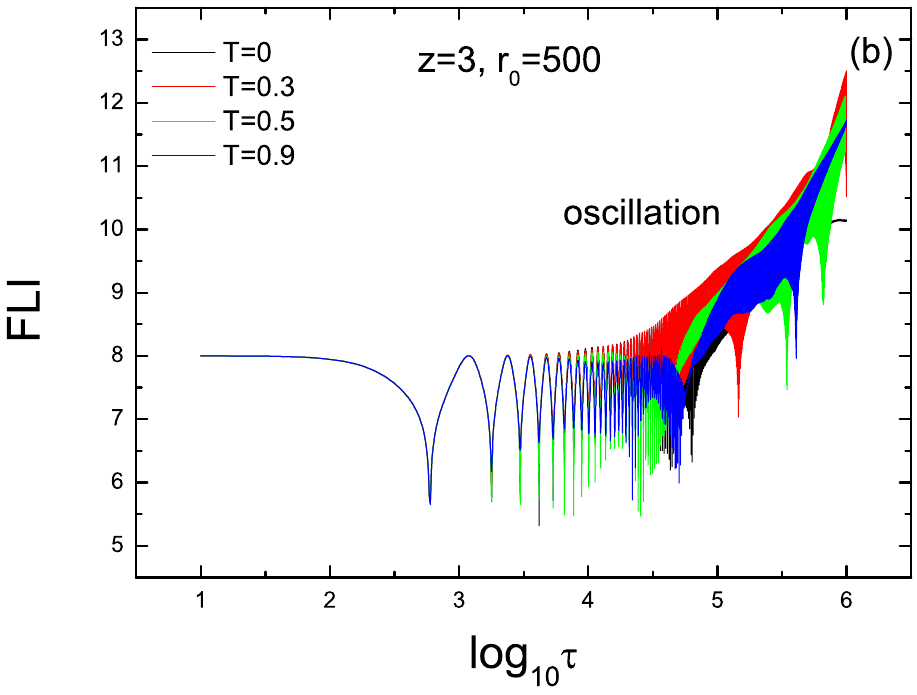}\\
    \includegraphics[scale=0.7, bb= 0 0 280 210]{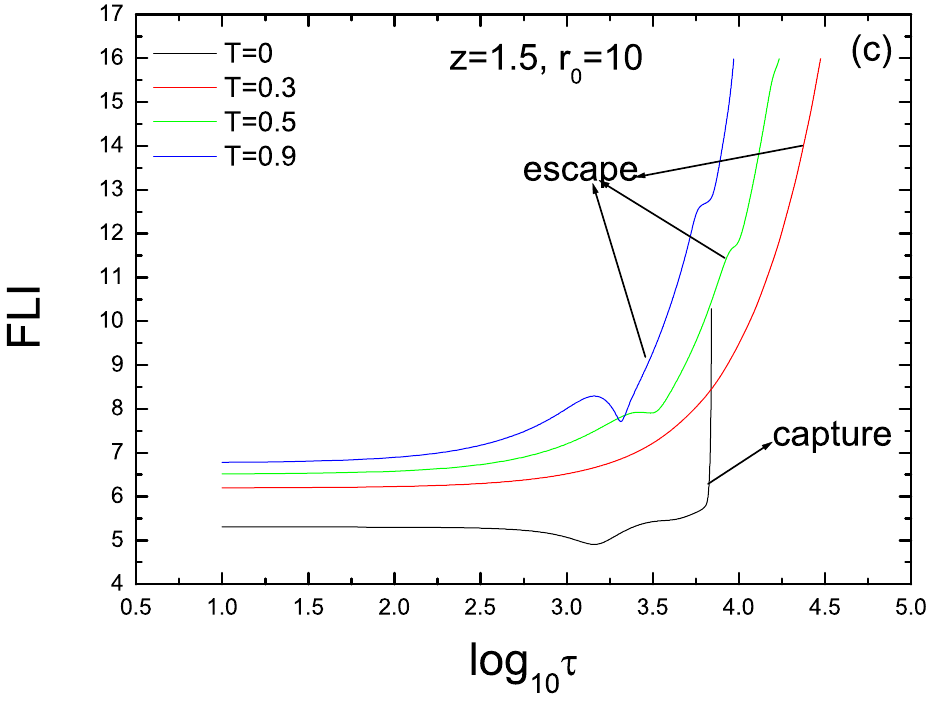}\ \hspace{0.8cm}
    \includegraphics[scale=0.7, bb= 0 0 280 210]{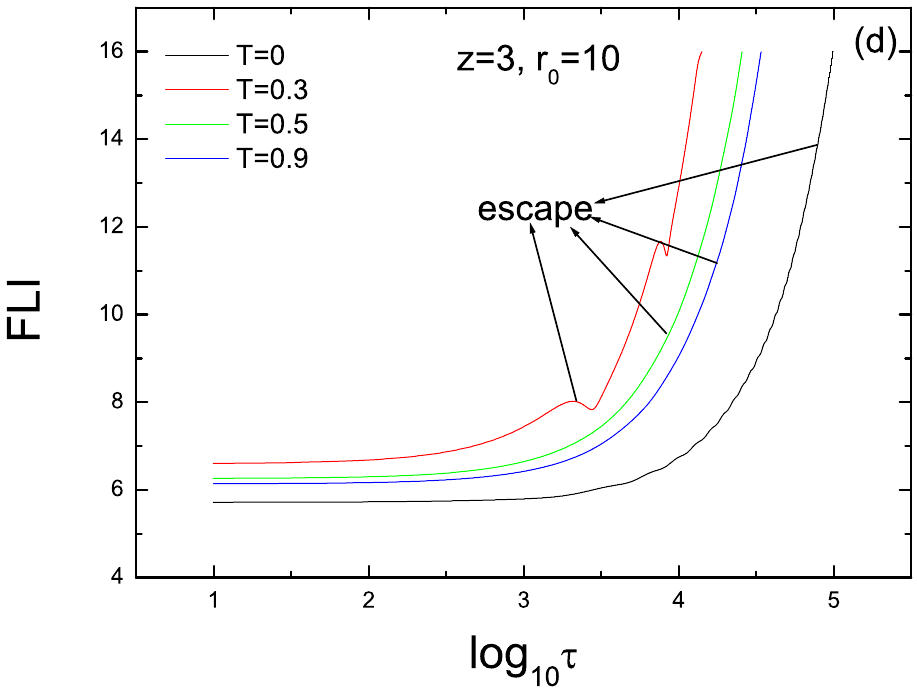}\\
    \caption{FLIs with different $z$ (left plots for $z=1.5$ and right plots for $z=3$) and $T$ for $r_{0}=500$ (plots above) and $r_{0}=10$ (plots below) with different $z$.
    }
    \label{fig_FLIs_dz_tem}
\end{figure}

Now we heat the system with Lifshitz dynamical exponent.
We firstly discuss the case of the initial position of string being large ($r_{0}=500$).
At zero temperature the string is captured by the black brane after a finite number of oscillations for $z<2$.
However, as the temperature is increased, the string begins to escape to infinity after a finite number of oscillations (see the above left plot in FIG.\ref{fig_FLIs_dz_tem},
in which $z=1.5$). While for $z=3$, the case is different. Even we heat the system, the string oscillates around the black brane till eternity as what happen at zero temperature
(see the above plot in FIG.\ref{fig_FLIs_dz_tem}). Therefore, for large Lifshitz dynamical exponent $z$, the motion of string is quasi-periodic even if
we heat the system.

Subsequently, we turn to discuss the case of the initial position of string is close to the black brane ($r_0=10$).
We find that for small Lifshitz dynamical exponent $z$ ($z=1.5$ we study here), the string rapidly escapes to infinity when we heat the system
(see bottom left plot in FIG.\ref{fig_FLIs_dz_tem}). It is different to what happen at zero temperature, at which the string rapidly falls into the black brane.
For large Lifshitz dynamical exponent $z$, as mentioned above, the string rapidly escapes to infinity at zero temperature.
Here we find that even if we heat the system, the string also rapidly escapes to infinity (see the below right plot in FIG.\ref{fig_FLIs_dz_tem}).

\subsubsection{The effect from HV exponent}

\begin{figure}
	\includegraphics[scale=0.7, bb= 0 0 280 210]{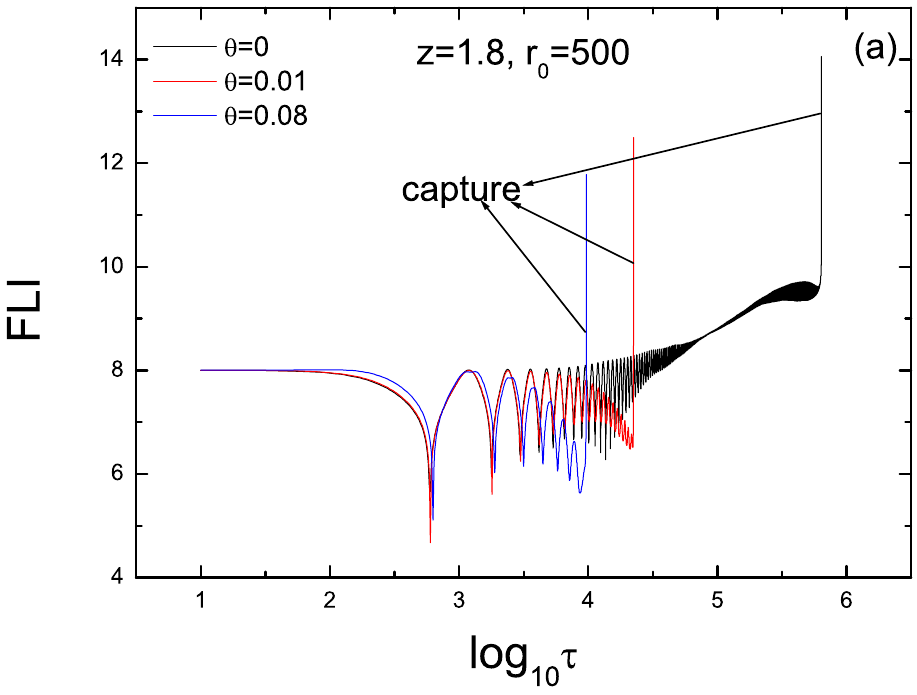}\ \hspace{0.8cm}
    \includegraphics[scale=0.7, bb= 0 0 280 210]{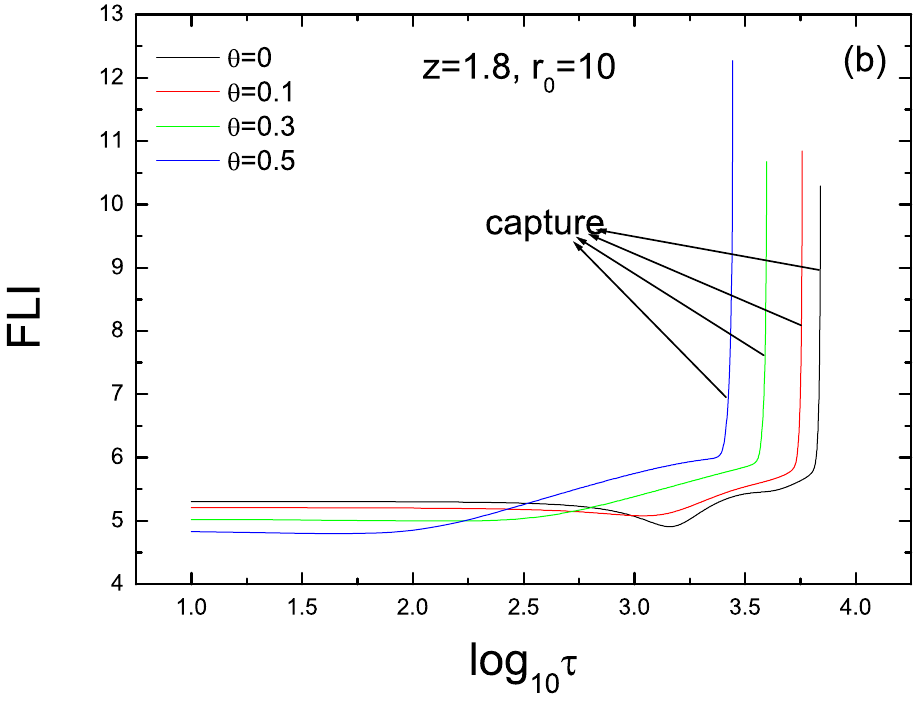}\\
	\includegraphics[scale=0.7, bb= 0 0 280 210]{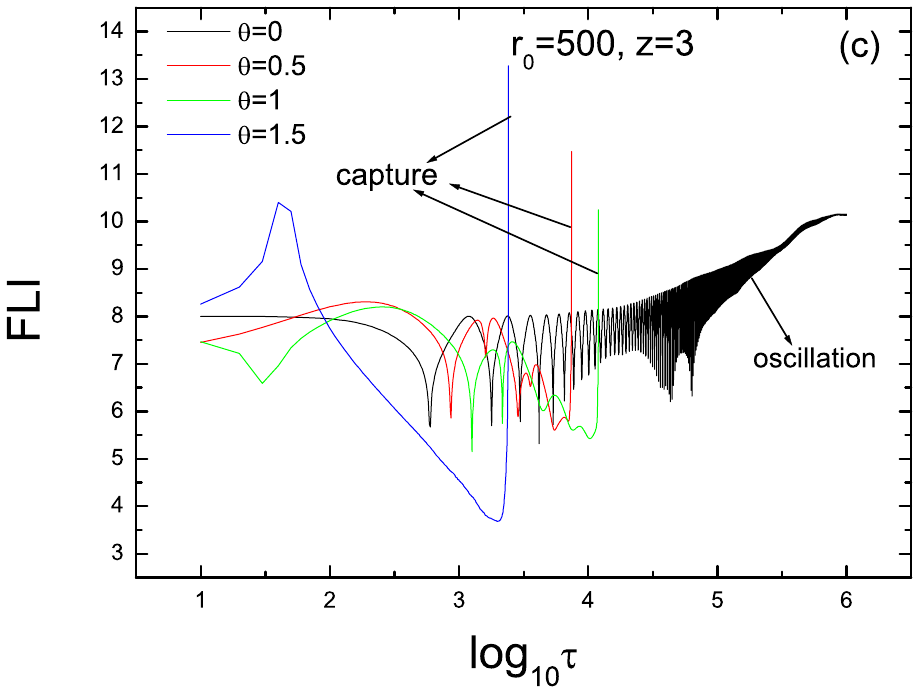}\ \hspace{0.8cm}
    \includegraphics[scale=0.7, bb= 0 0 280 210]{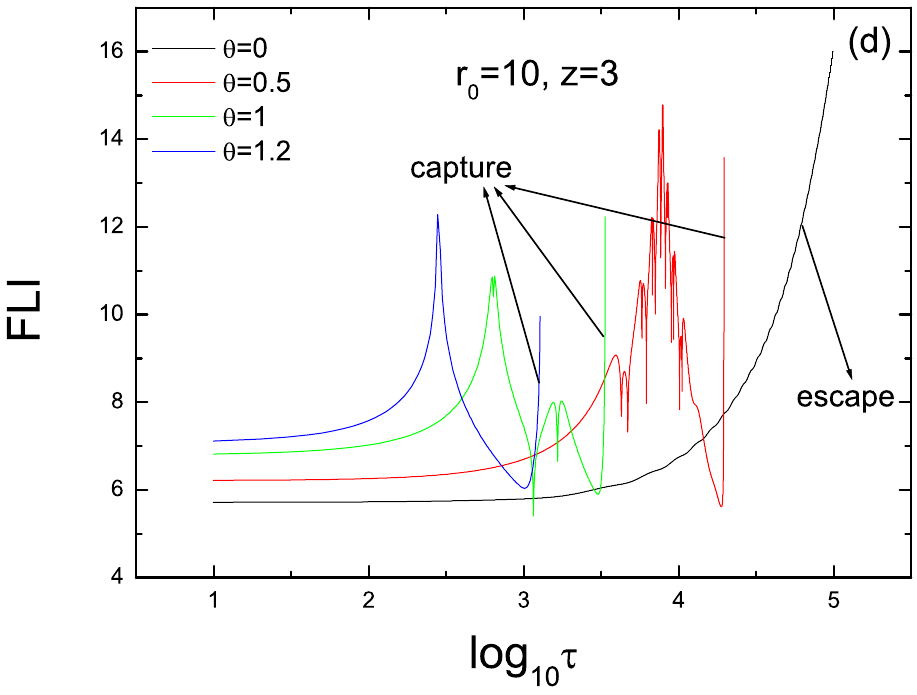}\\
    \caption{FLIs for $r_{0}=500$ (left plots) and $r_{0}=10$ (right plots) with different $z$ and $\theta$ at $T=0$.}
    \label{fig_theta_v1}
\end{figure}
%%%%%%%%%%%%%%%%%%%
\begin{figure}
	\includegraphics[scale=0.7, bb= 0 0 280 210]{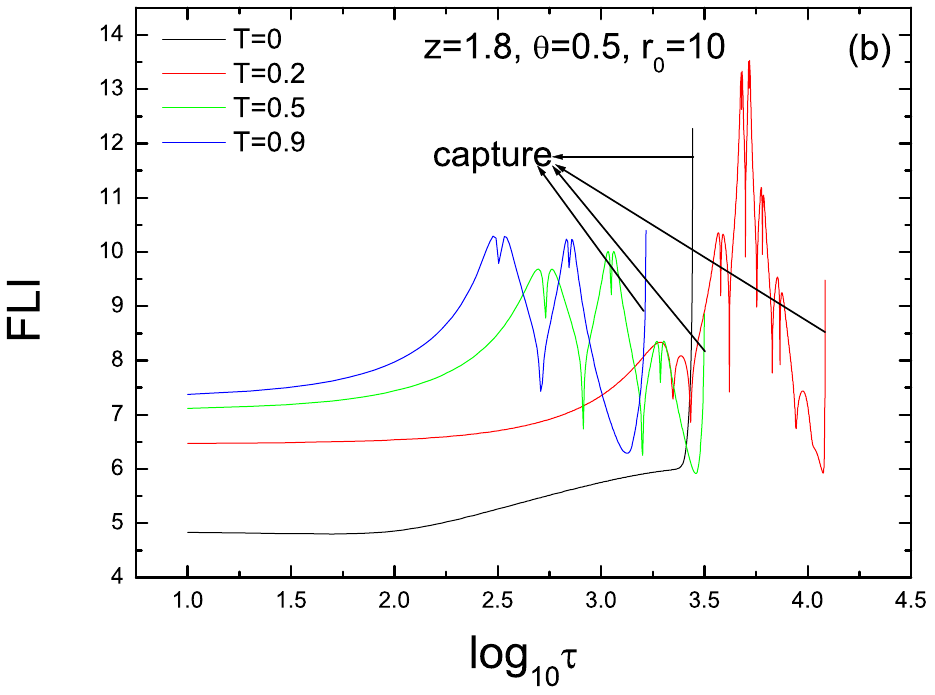}\ \hspace{0.8cm}
    \includegraphics[scale=0.7, bb= 0 0 280 210]{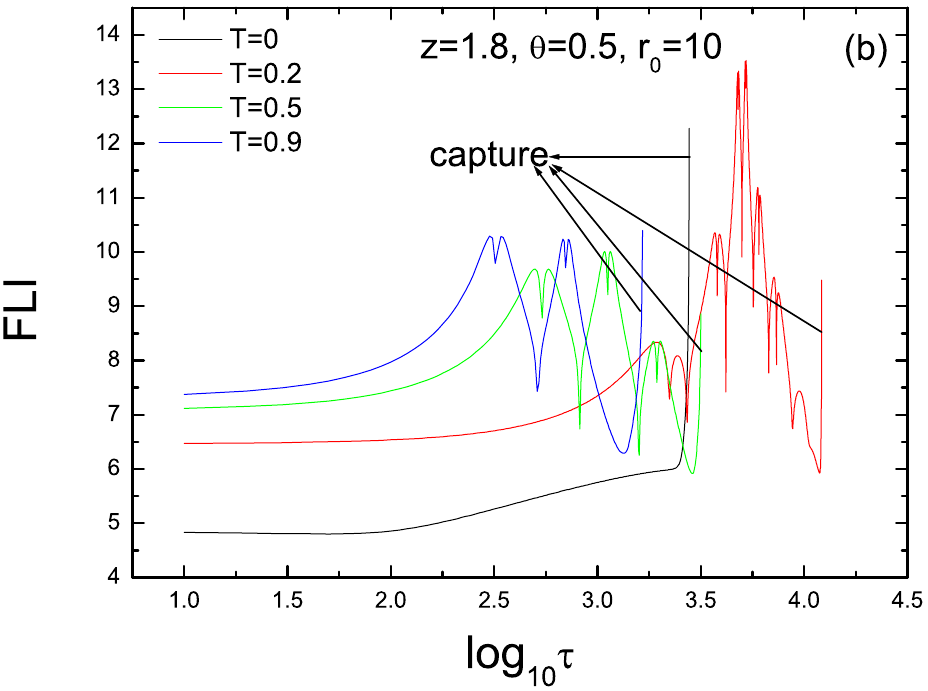}\\
    \includegraphics[scale=0.7, bb= 0 0 280 210]{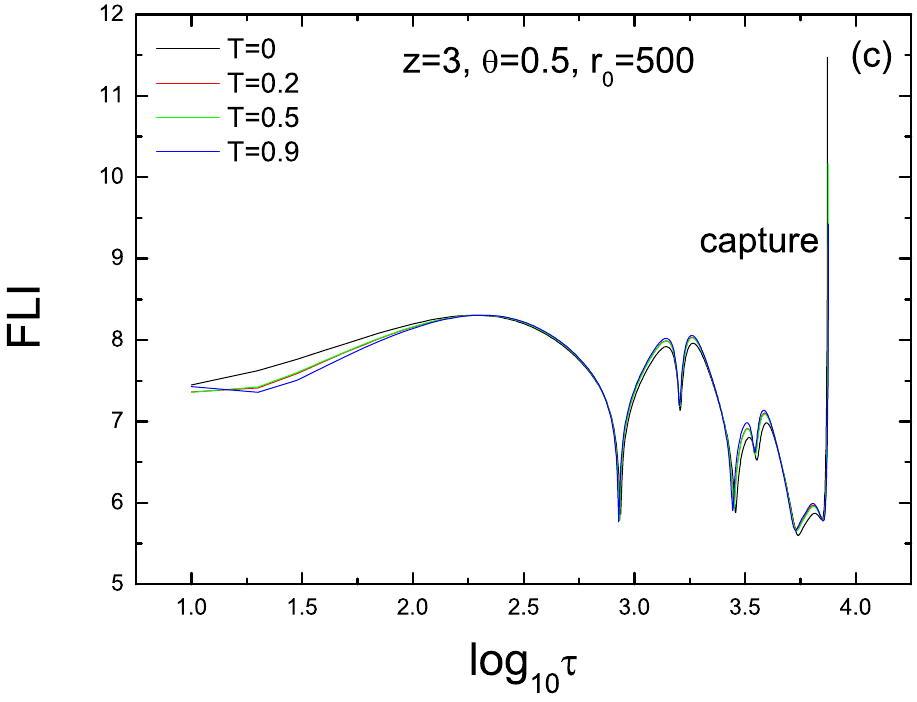}\ \hspace{0.8cm}
    \includegraphics[scale=0.7, bb= 0 0 280 210]{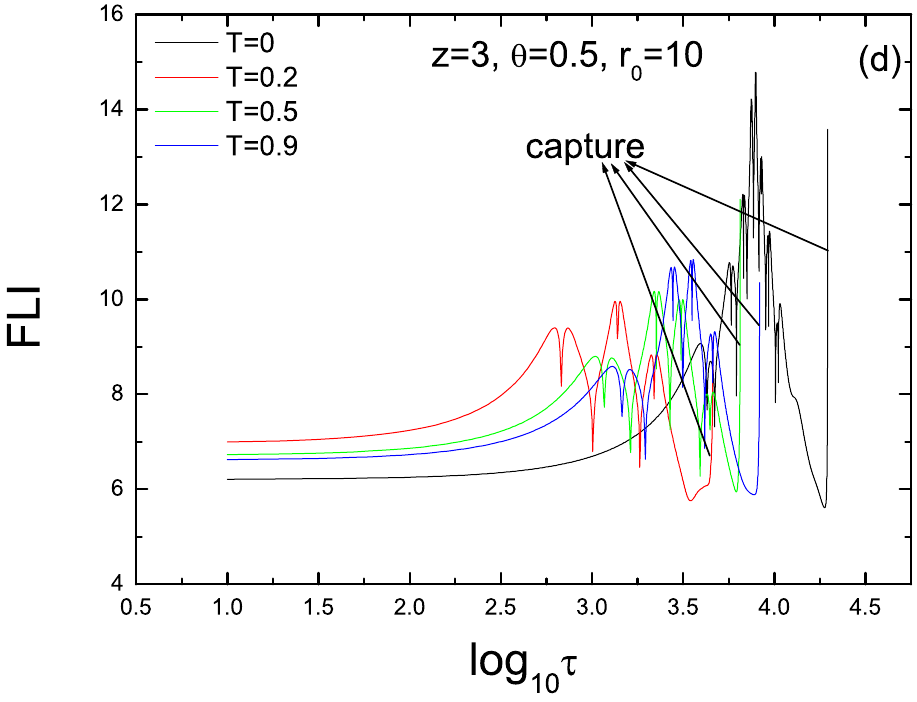}\\
    \caption{FLIs with different $z$, $\theta$ and $T$ for $r_0=500$ and $r_0=10$.
    }
    \label{fig_FLIs_dtheta_dtem_dz}
\end{figure}

In this subsection, we turn to explore the effect of HV exponent on the chaotic dynamics.
We first study the case of $z<2$.
For the initial position of the string locating far away from the black brane,
with the increase of $\theta$, the oscillating time before the string being captured becomes shorter (see the above left plot in FIG.\ref{fig_theta_v1}).
Especially, we find that for $z=1.8$ the string rapidly falls into black brane without any oscillating and the oscillating time is very short once the HV exponent $\theta$ is beyond the value of $\theta=0.1$.
When the initial position of the string is near the black brane, the string falls into the black brane for all allowed $\theta$ (see the above right plot in FIG.\ref{fig_theta_v1}).
With the increase of $\theta$, the falling velocity becomes faster.
Then we focus the case of $z>2$.
When the initial position of the string locates far away from the black brane,
with the increase of $\theta$, the string changes from the state of oscillating at the last stage to the state of captured by the black brane (see the left below plot in FIG.\ref{fig_theta_v1}).
When the initial position of the string is near the black brane, with the increase of $\theta$, the string falls into the black brane.
Note that for $\theta=0$, the string escapes to infinity.
At the same time, the falling velocity becomes faster as HV exponent $\theta$ becomes large.
In short, the HV exponent speeds up the falling of the string toward the black brane.

Further, we also heat the system with HV.
We find that the HV exponent plays a very important role in determining the state of the chaotic system.
From FIG.\ref{fig_FLIs_dtheta_dtem_dz}, we see that the chaotic system does not essentially changes when we heat the system.

\subsubsection{The effect of the winding number}

\begin{figure}
	\includegraphics[scale=0.7, bb= 0 0 280 210]{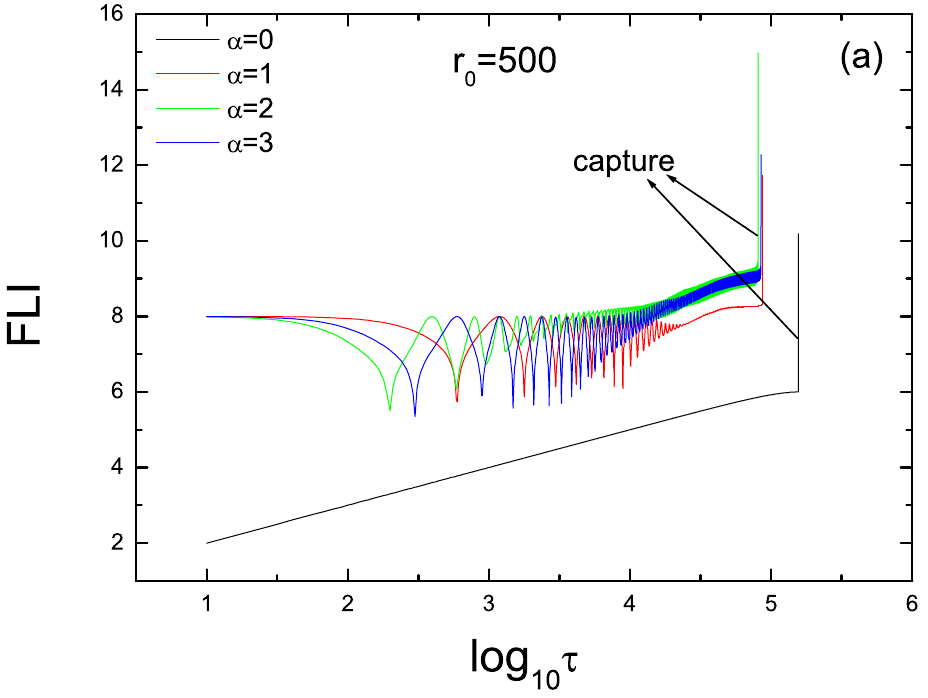}\ \hspace{0.8cm}
    \includegraphics[scale=0.7, bb= 0 0 280 210]{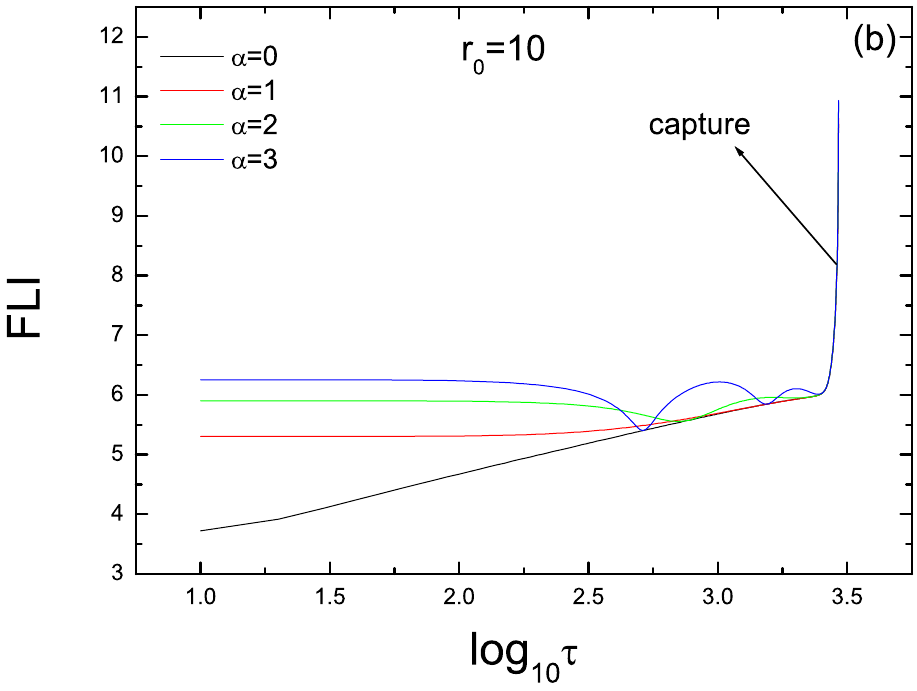}\ \\
    \caption{FLIs with different $\alpha$ for $r_{0}=500$ (left plot) and $r_{0}=10$ (right plot). Here, we set $z=1$, $T=0$.}
    \label{fig_dif_alpha}
\end{figure}

In this subsection, we briefly discuss the effect of the winding number $\alpha$.
FIG.\ref{fig_dif_alpha} shows FLIs with different $\alpha$.
When the initial position of string is far away from the black brane ($r_0=500$ here),
we find that for $\alpha>1$, the string first undergoes a long time oscillation and then falls into the black brane (left plot in FIG.\ref{fig_dif_alpha}).
While for $\alpha=0$, for which the closed string reduces to the particle,
the object don't undergo the oscillation but directly falls into the black brane (left plot in FIG.\ref{fig_dif_alpha}).
When the initial position of string is close to the black brane ($r_0=10$), the object falls into the black brane for all $\alpha$.
As by-product, we can see that when the object is free falling, the oscillating stage maybe the property of string motion.
It deserves future pursuit.

\begin{figure}
	\includegraphics[scale=0.7, bb= 0 0 280 210]{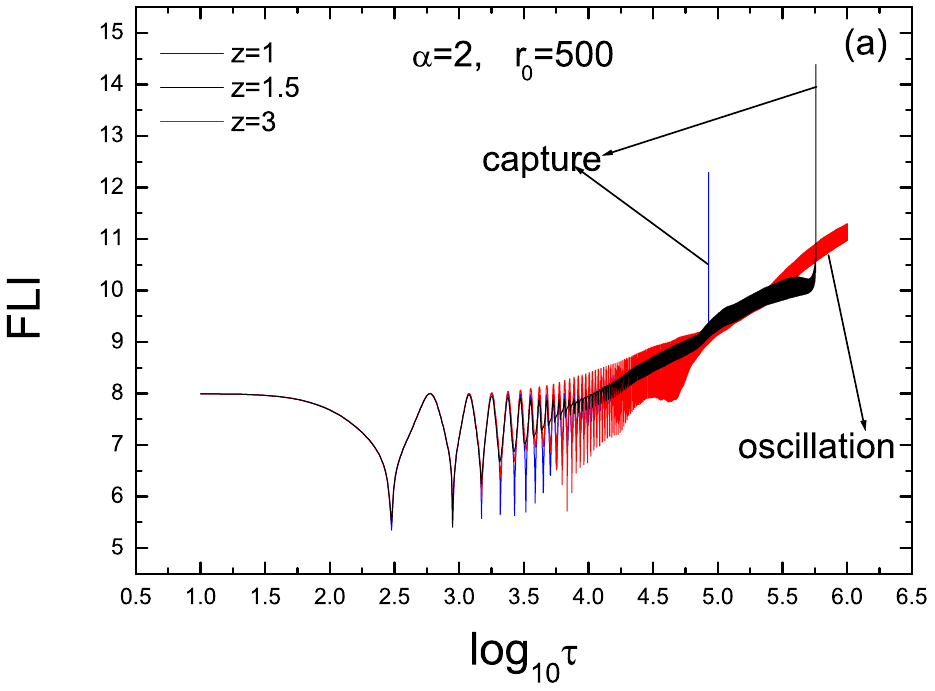}\ \hspace{0.8cm}
    \includegraphics[scale=0.7, bb= 0 0 280 210]{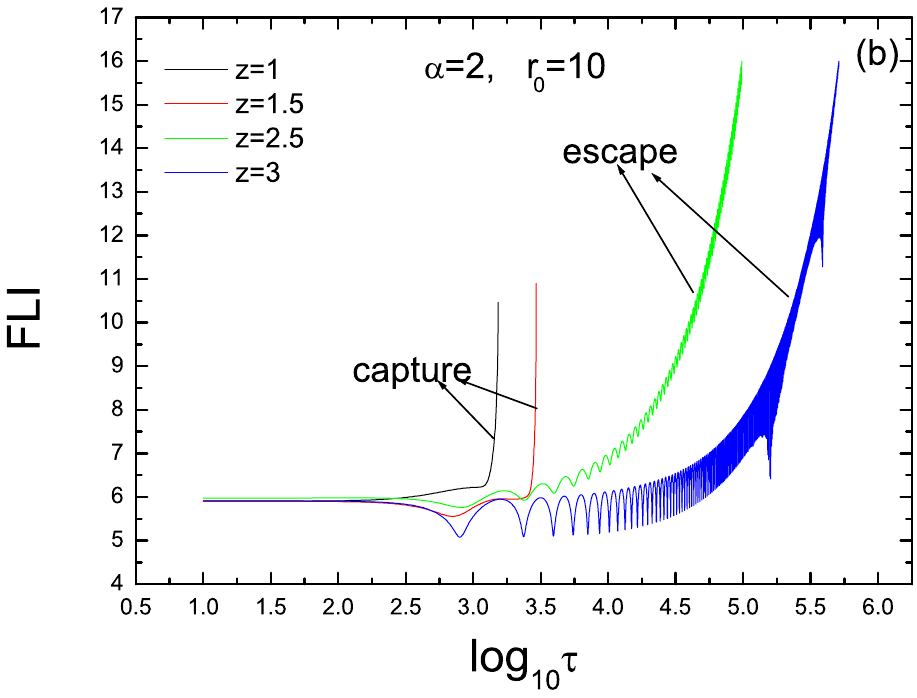}\ \\
    \caption{FLIs for $r_{0}=500$ and $r_{0}=10$ with different $z$. Here, $\alpha=2$, $T=0$. }
    \label{fig_alpha_2_z}
\end{figure}

\begin{figure}
	\includegraphics[scale=0.7, bb= 0 0 280 210]{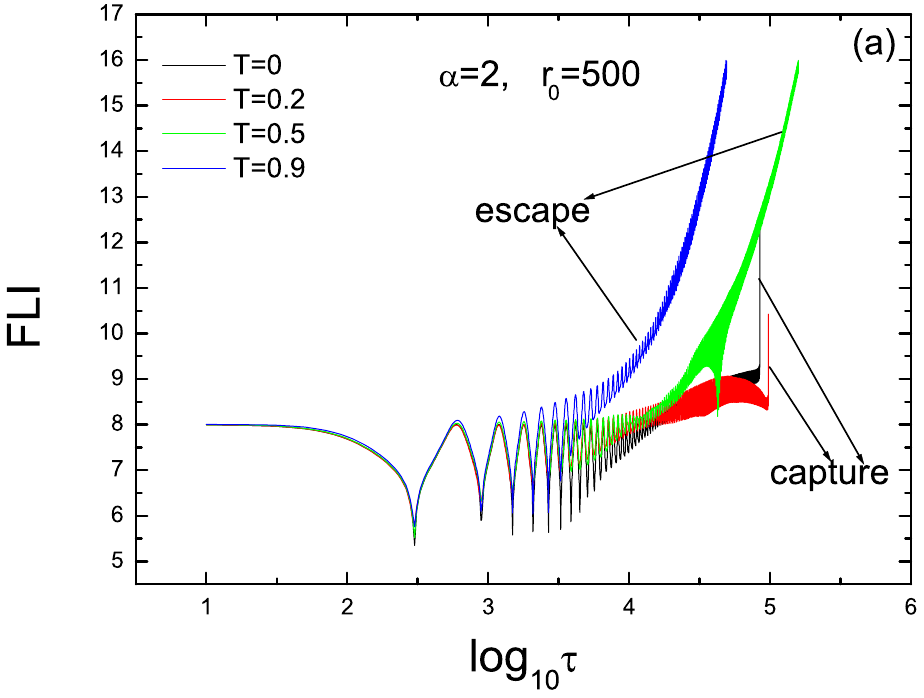}\ \hspace{0.8cm}
    \includegraphics[scale=0.7, bb= 0 0 280 210]{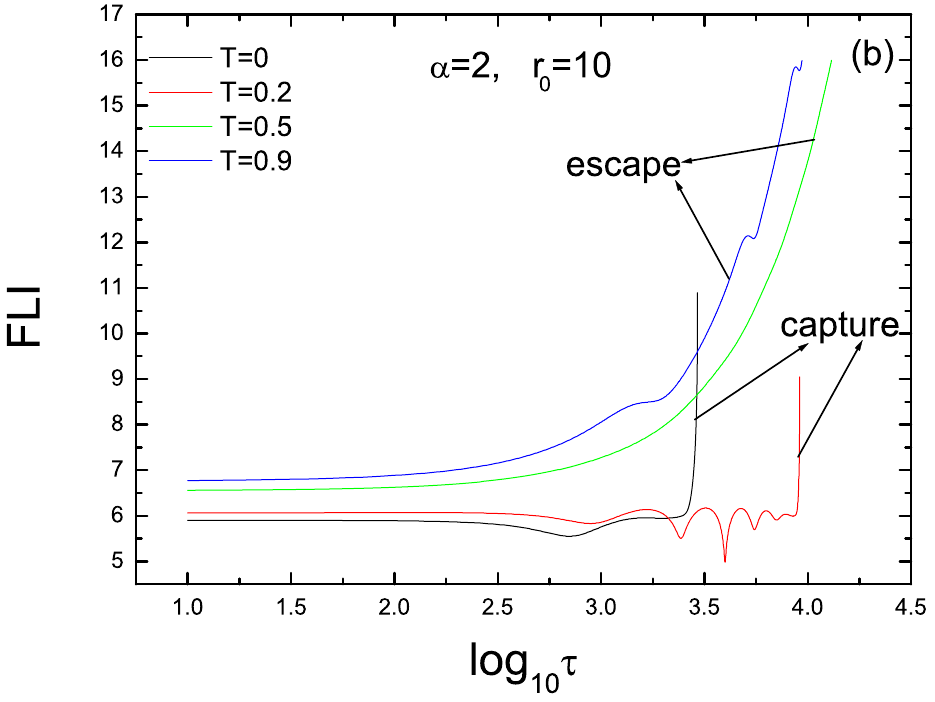}\ \\
    \caption{FLIs with the winding number of closed string $\alpha=2$ for different temperatures $T$.
    Left plot is for $r_{0}=500$ and right plot is for$r_{0}=10$. Here, we set $z=1$.}
    \label{fig_alpha_2_tem}
\end{figure}

Therefore, we conclude that the chaotic dynamics of the string seems to be insensitive to the winding number.
To confirm this viewpoint, we also show the FLIs of sting with different Lifshitz dynamical exponent $z$ (FIG.\ref{fig_alpha_2_z})
and different temperature $T$ (FIG.\ref{fig_alpha_2_tem}). Note that we fix $\alpha=2$.
We see that the results are not significant difference from the case of $\alpha=1$ (see FIG.\ref{fig_FLIs_ze1} for different temperature $T$
and FIG.\ref{fig_FLIs_dz} for different $z$ at zero temperature).

%%%%%%%%%%%%%%%%%%%%%%%%%
%%%%%%%%%%%%%%%%%%%

%\section{Some remarks on chaos bound and classical chaos in holography}

\section{Conclusions and discussions}\label{sec-con}

In this paper, we study the chaotic dynamics of closed string around charged HV black brane.
The Hawking temperature, Lifshitz dynamical exponent and HV exponent together affect the chaotic dynamics of this system.
The properties of chaotic dynamics are summarized as what follows.
\begin{itemize}
  \item \textbf{Extremal black brane}
  \begin{itemize}
    \item For zero HV exponent, i.e., $\theta=0$, there is a threshold value $z_{\ast}=2$, below which the string is captured by the black brane no matter where the string is placed.
  However, when $z>2$, the string escapes to infinity if it is placed near the black brane at the beginning,
  but if the initial position of string is far away from the black brane, it oscillates around the black brane till eternity,
  which is a quasi-periodic motion.
    \item HV exponent plays the role driving the string falling into the black brane.
    With the increase of $\theta$, the falling velocity becomes faster.
  \end{itemize}
  \item The chaotic dynamics of the string is insensitive to the winding number.
  \item When we heat the system, the string tends to escape to infinity.
  But there is an exception for $z>2$ that when the string is placed far away from the black brane,
  the string oscillates till eternity even if we heat the system.
  \item The HV exponent plays a very important role in determining the state of the chaotic system.
The chaotic system does not essentially changes when we heat the system with large HV exponent.
\end{itemize}

There are many questions deserving further exploration in future.
\begin{itemize}
  \item An immediate and important issue is the dual interpretation of the closed string.
  In ~\cite{Susskind:2018tei}, Susskind proposes that
  the momentum of the particle falling toward the black hole in the bulk corresponds to
  the operator growth in the chaotic quantum systems.
  We can follow this idea to explore the dual interpretation of the closed string.
  \item Further, we can also study the chaotic dynamics of closed string in the black brane background,
  which is dual to the momentum dissipation system, for example,
  the systems dual to Q-lattice geometry ~\cite{Donos:2013eha} or the Einstein-Maxwell-axions theory ~\cite{Andrade:2013gsa} or the massive gravity ~\cite{Vegh:2013sk}.
  These generalized explorations surely give further understanding on the role the momentum dissipation plays in the dual system.
  Also such studies maybe provide some clues to the interpretation of the closed string in the dual system.
  \item We can also extend this study to the particles, for example the test scalar particle ~\cite{Wang:2016wcj} or the fermions.
  \item It is interesting to study the relation among the chaos bound proposed in ~\cite{Maldacena:2015waa}, the black hole horizon and the string motion following
  ~\cite{Hashimoto:2016dfz,Dalui:2018qqv,Dalui:2019esx}.
\end{itemize}

\begin{acknowledgments}

This work is supported by the Natural Science Foundation of China under
Grant Nos. 11775036, 11703005, 11263003.
Z. D. Ma is also supported by the PhD start fund's MY2018B015
and J. P. Wu is also supported by Top Talent Support Program from Yangzhou University.

\end{acknowledgments}

%%%%%%%%%%%%%%
%\begin{appendix}

%\section{Numerical method}\label{app-num}

%\end{appendix}
%%%%%%%%%%%%%%%%%%%%%

%\end{CJK*}{GBK}{kai}
\end{document}